\documentclass[10pt]{article}
\usepackage{graphicx,xcolor,amssymb}
\usepackage[titletoc,toc,title]{appendix}
\usepackage{rotating}
\usepackage{pdflscape}
\usepackage{array}
\usepackage{amsmath}

\usepackage{bm}
\usepackage{lipsum}
\usepackage{listings}
\usepackage{caption}
\definecolor{mygreen}{rgb}{0,0.6,0}
\definecolor{mygray}{rgb}{0.5,0.5,0.5}
\definecolor{mymauve}{rgb}{0.58,0,0.82}
\definecolor{gris245}{RGB}{245,245,245}
\definecolor{olive}{RGB}{50,140,50}
\definecolor{brun}{RGB}{175,100,80}
\definecolor{dkgreen}{rgb}{0,0.6,0}
\definecolor{mauve}{rgb}{0.58,0,0.82}
\definecolor{black}{rgb}{0.0, 0.0, 0.0}
\definecolor{beaver}{rgb}{0.62, 0.51, 0.44}
\definecolor{columbiablue}{rgb}{0.61, 0.87, 1.0}
\definecolor{lightpink}{rgb}{1.0, 0.71, 0.76}
\definecolor{amber}{rgb}{1.0, 0.75, 0.0}
\definecolor{amber(sae/ece)}{rgb}{1.0, 0.49, 0.0}
\definecolor{aquamarine}{rgb}{0.5, 1.0, 0.83}
\definecolor{cadmiumred}{rgb}{0.89, 0.0, 0.13}
\definecolor{cambridgeblue}{rgb}{0.64, 0.76, 0.68}
\definecolor{electricblue}{rgb}{0.49, 0.98, 1.0}
\definecolor{paleplum}{rgb}{0.8, 0.6, 0.8}

\usepackage{pgffor} 
\usepackage{forloop} 
\usepackage{hyperref}
\hypersetup{
  colorlinks=true,
  linkcolor=blue,
  filecolor=magenta,      
  urlcolor=cyan}

\usepackage{tikz}
\usetikzlibrary{arrows.meta}

\begin{document}
\title{Thomson problem on a spherical cap}
\author{Paolo Amore \\
\small Facultad de Ciencias, CUICBAS, Universidad de Colima,\\
\small Bernal D\'{i}az del Castillo 340, Colima, Colima, Mexico \\
\small paolo@ucol.mx}

\maketitle

\begin{abstract}
We investigate the low-energy configurations of $N$ mutually repelling charges confined to a spherical cap and interacting via the Coulomb potential. In the continuum limit, this problem was solved by Lord Kelvin, who found a non-uniform charge distribution with an integrable singularity at the boundary. To explore the discrete analogue, we developed an efficient numerical method that enables energy minimization while maintaining the number of charges at the cap’s edge fixed. Using this approach we have obtained numerical results for various values of $N$ and cap angular widths. Based on these results, we analyze the emergence and behavior of topological defects as functions of both $N$ and the cap’s curvature.
\end{abstract}

\section{Introduction}
\label{sec:intro}

The problem considered in the present paper is a variant of the well-known Thomson problem, which consists in finding the ground state of a system of $N$ identical charges that repel each other and are constrained to lie on the surface of a sphere. This problem was first introduced by J.J. Thomson~\cite{Thomson1904} more than a century ago as a model of the atom, but it was later abandoned with the advent of quantum mechanics.

Even though the original motivations for studying it have since disappeared, it was  rediscovered in more recent times in refs.~\cite{Erber91, Edmundson92, Glasser92, Erber95, Erber97} as an example of classical many body system with a complex behavior. In particular Erber and Hockney were the first to point out that the number of local minima for this problem grows exponentially with $N$ and finding the global minimum of the total energy becomes very challenging when $N$ is sufficiently large.  Indeed, the study of large configurations requires the use of sophisticated algorithms, that explore to some extent the energy landscape of the problem (see for instance refs.~\cite{Altschuler94,Morris96,Altschuler97,Xiang97,Wales06,Wales09,Lai24}). 
Numerical experiments on configurations of different sizes have shown that the nature of the local minima changes as $N$ grows, evolving from configurations with isolated pentagonal disclinations to more complex arrangements of topological defects (such as the rosette observed in \cite{Wales06,Wales09}). The appearance of these topological defects on the sphere is a consequence of Euler's theorem of topology, which constrains the total topological charge to be $Q_{\rm total} = 12$, without restricting the number of individual defects. 
 
While it is not possible to study this problem analytically except for very small systems~\cite{Schwartz13,Schwartz20}, at least two different lines of research can be identified in the limit $N \rightarrow \infty$. On one hand, Bowick and collaborators~\cite{Bowick00,Bowick02,Bowick09} have developed a formalism in which the fundamental degrees of freedom are the topological defects rather than the $N$ charges themselves. On the other, Saff and collaborators~\cite{Saff94b,Saff97,Saff19} have studied the asymptotic behavior of the energy for systems of particles interacting via potentials of the form $1/r^s$, where $r$ is the Euclidean distance between two particles and $s \geq -1$ is a real exponent ($s=1$ corresponds to the Coulomb potential). For sufficiently large $N$, the behavior of discrete systems can be compared with the descriptions obtained in these two approaches.

The occurrence of topological defects is also observed in domains that are flat but have a border, due to the geometrical frustration introduced by the boundary (in this case, however, Euler's theorem constrains the total topological charge to $Q_{\rm total } = 6$). An example of this is the flat disk studied in refs.~\cite{Bedanov94,Bedanov95,Koulakov98,Oymak01,Peeters03a,Peeters03b,Worley06,Moore07,Olvera13,Cerkaski15,Cerkaski17a,Cerkaski17b,Amore17, Amore23}. The defects observed in this system differ noticeably from those on the sphere, with the border playing an important role. At large $N$, the border carries a large positive topological charge due to a typical arrangement of alternating pentagonal and square cells (see Fig.14 of \cite{Amore23}). This is followed by a layer of cells containing many heptagons (and occasionally octagons), which rapidly shields the topological charge (see also Fig.9 of \cite{Amore23}). Internal defects, on the other hand, tend to form long sequences of alternating pentagonal–heptagonal cells, creating large filaments at large $N$ (see Fig.~17 of \cite{Amore23}).

The problems described above can be regarded as extreme cases: a domain that is curved but has no border (the sphere), and a domain that is flat but with a border (the disk). The mechanisms that generate topological defects in the two cases are different, and this is reflected in the nature and distribution of the defects themselves.

In this paper we study the Thomson problem on a spherical cap of fixed angular width $\theta_{\rm max}$, as an example of a system with both curvature and a border. This geometry naturally interpolates between the two problems described above: for $\theta_{\max} \rightarrow 0^+$, while keeping the surface area fixed, it reduces to the Thomson problem on a disk, whereas for $\theta_{\max} \rightarrow \pi^-$ it reduces to the standard Thomson problem on the sphere. The relative importance of curvature and boundary effects can be tuned simply by varying $\theta_{\rm max}$.

Although this variant of Thomson problem has never been  studied before,  the problem of packing on a spherical cap has recently been studied in ref.~\cite{Amore24}; 
moreover, Azadi and Grason~\cite{Grason14} have studied a continuum elasticity model of crystalline caps bound to a spherical substrate and found that  the ground state for this problem has a characteristic behavior with radially oriented defect scars emanating from the border and terminating in the bulk: interestingly, in our simulations we have found several configurations with this behavior, for caps of small curvature (particularly for $\theta_{\rm max} = 0.1 \pi$); Li et al.~\cite{Grason19} have also investigated the transition between defect-free caps to single-disclination ground states for caps of different width.

The paper is organized as follows: Section~\ref{sec:continuum} describes the continuum problem using the solution obtained long ago by Lord Kelvin~\cite{Kelvin47}; Section~\ref{sec:discrete} presents the discrete model of $N$ charges on a cap and the numerical approach used to calculate low-energy solutions; Section~\ref{sec:results} reports the numerical results for various values of $N$ and $\theta_{\rm max}$; finally, Section~\ref{sec:concl} summarizes our conclusions.

\section{Continuum limit}
\label{sec:continuum}

We consider a conducting spherical bowl (cap) of angular width $\theta_{max}$ and total charge $Q$: the total electrostatic energy can be expressed as
\begin{equation}
\mathcal{E} =  \frac{1}{2} \int \int  \frac{  \sigma({\bf r})  \sigma({\bf r}')  }{|{\bf r}- {\bf r}'|} \  d^2r  d^2{r'}   \  ,
\end{equation}
where $\sigma({\bf r})$ is the surface charge density and
\begin{equation}
Q = \int \sigma({\bf r}) d^2r   
\end{equation}
is the total charge  (because we are interested in comparing the continuum distribution with a discrete distribution with the same charge, in the following we will 
assume $Q=N$, where $N$ is the number of unit charges on the cap).

Adapting the discussion of Mughal and Moore~\cite{Moore07}  for the disk to the spherical cap, we introduce  a Lagrange multiplier $\mu$ and write the constrained equation for the energy as
\begin{equation}
\mathcal{E} =  \int  \sigma({\bf r}) \left[\frac{1}{2}  \int \left(\frac{\sigma({\bf r}')}{|{\bf r}-{\bf r}'|}-\mu \right)  d^2r' \right] d^2r \ .
\end{equation}

From the condition  $\delta E \equiv E(\sigma + \delta\sigma) -E(\sigma)=0$ it follows that
\begin{equation}
\mu  =  \int d^2r' \frac{\sigma({\bf r}')}{|{\bf r}-{\bf r}'|}  \ .
\label{eq_Fredholm}
\end{equation}

This in fact is an integral equation for the charge density $\sigma$ (eqs.(9) and (10) of ref.~\cite{Moore07} report the solution to the equation above for the case of the disk) and the total energy takes the form
\begin{equation}
\mathcal{E} = \frac{\mu N}{2} \ .
\end{equation}

\begin{figure}[t]
  \centering
\begin{tikzpicture}[
    >=Stealth,
    dot/.style={circle, fill, inner sep=1.5pt}
]
  \def\R{3}
  
  \draw[thick] (210:\R) arc[start angle=210, end angle=-30, radius=\R];
  
  \coordinate (O) at (0,0);
  \node[dot, label={below right:$$}] at (O) {};
  
  \coordinate (M) at (90:\R);
  \node[dot, label={above:$$}] at (M) {};
  
  \coordinate (Eleft)  at (210:\R);
  \coordinate (Eright) at (-30:\R);
  

  \draw[thin] (M) -- (Eright)
    node[midway, right] {$a$};

  \coordinate (Fpoint) at (150:\R);
  \draw[thin] (Eright) -- (Fpoint)
    node[pos=0.6, below=5pt] {$f$};

  \draw[dashed] (O) -- (Eright);
  \draw[dashed] (O) -- (M);

  \begin{scope}
    \draw[->] (90:0.7) arc[start angle=90, end angle=-30, radius=0.7];
    \node at (30:1) {$\theta_{max}$};
  \end{scope}

\end{tikzpicture}
\caption{Spherical cap}
\label{fig:bowl}
\end{figure}
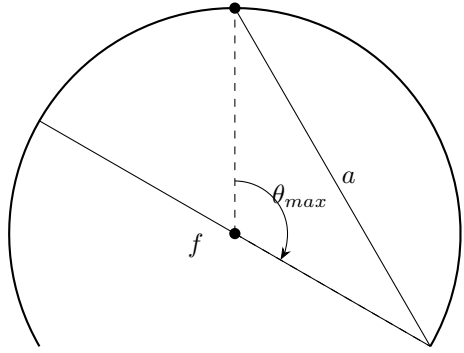

Luckily enough, Lord Kelvin~\cite{Kelvin47} calculated long time ago the charge density of a conducting spherical bowl; the  expression of the charge density can 
also be found  in ref. \cite{Maxwell92} at pag. 276,  ref.~\cite{Jeans33}, pag. 257 and ref.~\cite{McDonald02}  and reads
\begin{equation}
\begin{split}
\sigma_i  &= \frac{V}{2\pi^2 f} \left\{   \sqrt{\frac{f^2-a^2}{a^2-r^2}} -\arctan \sqrt{\frac{f^2-a^2}{a^2-r^2}}  \right\} \\
\sigma_o &=  \sigma_i + \frac{V}{2\pi a} \\
\end{split} \ ,
\end{equation}
where, using the notation of ref.~\cite{Kelvin47},  $V$ is the electrostatic potential of the bowl, $f$ is the diameter of the sphere containing the bowl, $r$ is the distance of a point on the bowl from the middle point of the bowl and $a$ is the maximum value that $r$ can take.  $\sigma_i$ and $\sigma_o$ are the surface charge density on the outside and on the inside of the cap respectively.
The configuration is shown schematically in Fig.~\ref{fig:bowl}, which  is an adaptation of Fig.80 of ref.~\cite{Jeans33}.

Using spherical coordinates  we have
\begin{equation}
\begin{split}
a = f \sin \left(\frac{\theta _{max}}{2}\right) \hspace{1cm} , \hspace{1cm}
r = f \sin \left(\frac{\theta }{2}\right)  \ ,
\end{split}
\end{equation}
and we can express the density in terms of the $\theta_{\max}$ as 
\begin{equation}
\begin{split}
\sigma_i(\theta) &=  \frac{V}{4 \pi ^2 R} \left(\sqrt{\frac{\cos \left(\theta _{\rm max }\right)+1}{\cos (\theta )-\cos \left(\theta _{\rm max}\right)}}  \right. \\
   &- \left. \tan ^{-1}\left(\sqrt{\frac{\cos \left(\theta _{\rm max }\right)+1}{\cos (\theta )-\cos\left(\theta _{\rm max }\right)}}\right)\right) \\
\sigma_o(\theta) &=   \sigma_i(\theta) +  \frac{V}{4 \pi  R\left(\theta _{\rm max }\right)}
\end{split}  \  .
\label{eq_sigmas}
\end{equation}

For our purposes we find convenient to consider a spherical cap of fixed surface $A = \pi$, corresponding to the area of a unit disk. In such case  we have
\begin{equation}
f = 2 R = \csc \left(\frac{\theta _{max}}{2}\right) \ .
\label{eq_radius}
\end{equation}

Notice that the ratio $\sigma_o(\theta)/\sigma_i(\theta)$ blows up for $\theta_{\rm max} \rightarrow \pi^{-}$ as
\begin{equation}
\begin{split}
\sigma_o(\theta)/\sigma_i(\theta) &\approx \frac{24 \pi  \left| \cos\left(\frac{\theta }{2}\right)\right|^3}{\left(\pi -\theta _{max}\right)^3} \\
&+\frac{3 \pi  \cos \left(\frac{\theta }{2}\right) (5 \cos (\theta )-13)}{10 \left(\pi -\theta _{\max
   }\right)} +1 + \dots
\end{split}
\end{equation}
whereas for $\theta_{\rm max} \rightarrow 0^+$:
\begin{equation}
\sigma_o(\theta)/\sigma_i(\theta) \approx 1 \ ,
\end{equation}
as expected from symmetry considerations.

As noted by Lord Kelvin: "The results for bowls of $270^0$ and $340^0$ illustrate the tendency of the whole charge to the convex surface, as the case of a thin spherical conducting surface with an infinitely small aperture is approached" (see ref.~\cite{Jeans33} at pag.251).

From the normalization of the total density, $\sigma(\theta) = \sigma_i(\theta) + \sigma_o(\theta)$,  we can express $V$ in terms of the total charge of the cap, $N$, as
 \begin{equation}
V =    \frac{\pi  N/R}{\left( \theta _{\max }+\sin \left(\theta _{\max }\right)\right)} \ .
 \end{equation}
 
If $\sigma(\theta)$ obtained from eqs.~(\ref{eq_sigmas}) is the correct solution then it must solve the Fredholm equation (\ref{eq_Fredholm}) and provide  $\mu$ independent of the
angles on the cap: we have numerically verified that this is the case for arbitrary $\theta$ and also exactly calculated explicitly the value of the integral for $\theta=0$, confirming that
the Lagrange multiplier $\mu$ is just the electrostatic potential on the cap, $\mu = V$.

Consequently the total electrostatic energy of a cap of radius $R$ is 
\begin{equation}
\mathcal{E}(\theta _{max}) =  \frac{\pi  N^2}{2 R \left( \theta _{\max }+\sin \left(\theta _{\max }\right)\right)}  \ .
\label{eq:Econtinuum}
\end{equation}

Using the value of $R$ obtained from eq.~(\ref{eq_radius}) we have
\begin{equation}
\left. \mathcal{E}(\theta _{max}) \right|_{fixed \ area}=\frac{\pi  N^2 \sin \left(\frac{\theta _{\max }}{2}\right)}{\theta _{\max }+\sin \left(\theta _{\max }\right)} \ .
\label{eq:Econtinuum2}
\end{equation}

Another possible choice is to keep the volume fixed, say to $v = \pi /2$ (volume of a sphere of radius $1/2$) and obtain
\begin{equation}
\left. \mathcal{E}(\theta _{max}) \right|_{fixed \ vol}  = \frac{\pi  N^2 \sin ^{\frac{2}{3}}\left(\frac{\theta _{\max }}{2}\right)}{\theta _{\max }+\sin
   \left(\theta _{\max }\right)} \ ,
\end{equation}
that diverges as $\theta_{\rm max}^{-1/3}$ for $\theta_{\rm max} \rightarrow 0$ and has a shallow minimum at $\theta_{\rm max} \approx 1.8955$.

\begin{figure}
\begin{center}
\includegraphics[width=\columnwidth,clip]{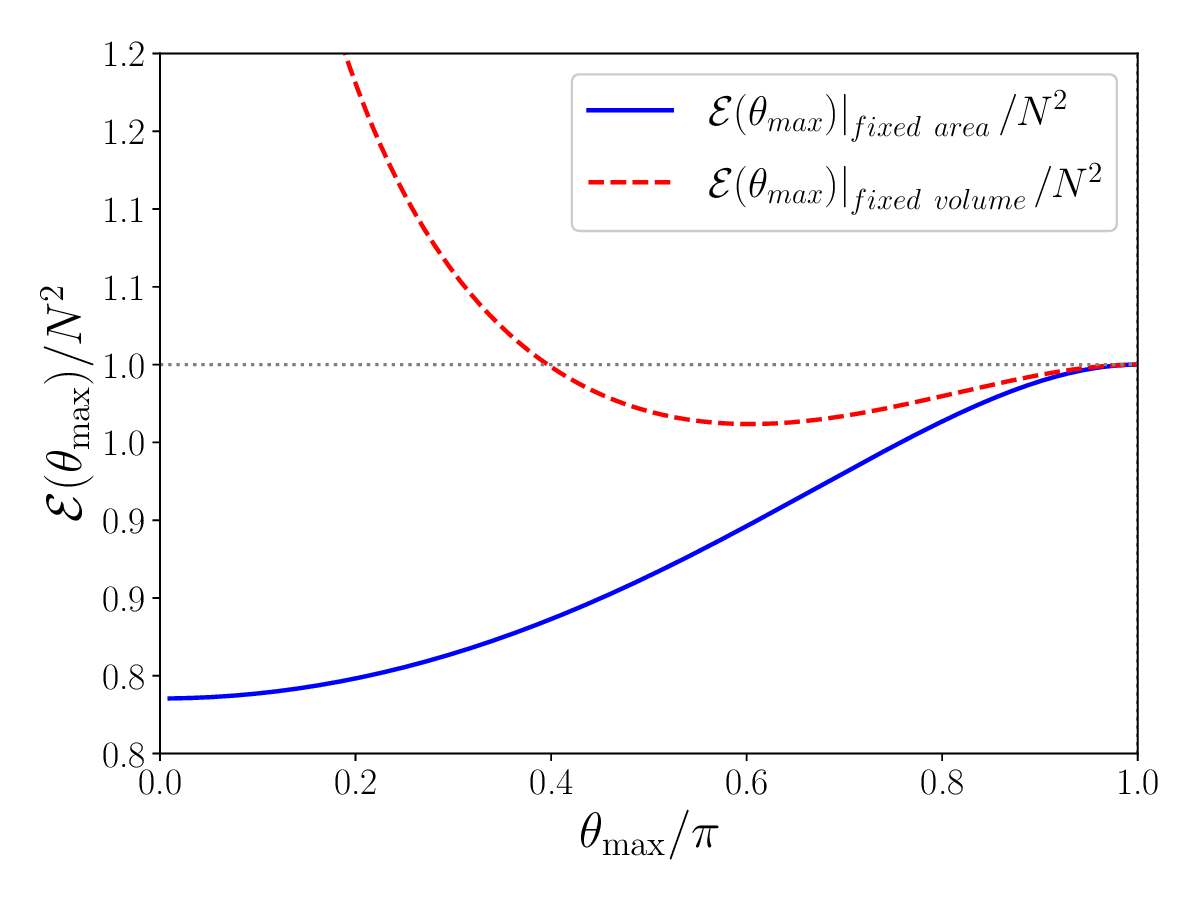}
\hspace{1cm}
\caption{Electrostatic energy of a conducting  spherical cap of angular width $\theta_{\rm max}$.}
\label{Fig_energy_cap}
\end{center}
\end{figure}

The two energies are plotted in Fig.~\ref{Fig_energy_cap}: notice that $\left. \mathcal{E}(\theta _{max}) \right|_{fixed \ area}$ 
correctly interpolates between the known energies of a unit disk ($\lim_{\theta _{max}\rightarrow 0 }\left. \mathcal{E}(\theta _{max}) \right|_{fixed \ area}  =\pi N^2/4$ ) and of a sphere of radius $R(\pi) = \frac{1}{2}$ ($\left. \mathcal{E}(\pi) \right|_{fixed \ area}  = \frac{N^2}{2 R(\pi)} = N^2$).

Taking inspiration from the approaches used by Worley~\cite{Worley06} and Mughal and Moore~\cite{Moore07} for  the problem of the conducting disk, we can use a qualitative argument to obtain  the dependence of $N_b$ (the border charges in the discrete system) on the total number of charges, $N$.

Our argument goes as follows: imagine that $N_b$ charges are distributed on the border of the spherical cap and assume that their distribution is uniform: in this case the
angular separation between contiguous charges is $\delta(N_b) = 2\pi/N_b$. We may assume that the section of the cap containing the border charges 
corresponds to $\theta_{max} - \delta_\perp  < \theta < \theta _{max}$, where $\delta_\perp$ is the angular width for the border charges in the direction perpendicular to the border. It is reasonable to assume  that $\delta_\perp \propto \delta(N_b)$. In this way we have
\begin{equation}
N_b =  2\pi \int_{\theta_{max}-\delta_\perp(N_b)}^{\theta_{max}} \sigma(\theta) \sin\theta d\theta
\end{equation}
which is a transcendental  equation for $N_b$.

To solve this equation we assume that $\delta_\perp \approx \kappa \sin\theta_{\rm max} \  \delta$ and $N_b \approx \lambda N^\alpha$ for $N \gg 1$; under these conditions the equation becomes
\begin{equation}
\lambda  N^{\alpha } \approx  N^{1-\frac{\alpha }{2}} \frac{2 \kappa  \sin \left(\theta _{max}\right) \cos \left(\frac{\theta_{max}}{2}\right)}{ \sqrt{\pi \kappa  \lambda } \left(\theta _{max}+\sin
   \left(\theta _{max}\right)\right)} +\dots
\end{equation}
requiring $\alpha = \frac{2}{3}$,  which is the same coefficient found for the disk~\cite{Worley06,Moore07}. Notice that the coefficient of $N^{1-\frac{\alpha }{2}}$ vanishes for $\theta_{\rm max}\rightarrow \pi$, 
thus requiring $\lambda =0$ and hence $N_b=0$, a result consistent with the fact that the sphere has no border; for $\theta_{\rm max} \rightarrow 0$, on the other hand, one obtains $\lambda = \left(\kappa/\pi\right)^{1/3}$.

\section{Discrete model}
\label{sec:discrete}

We now discuss the discrete model: formally this amounts to consider a density
\begin{equation}
\sigma({\bf r}) = \sum_{i=1}^N \delta( {\bf r} -{\bf r}_i)
\end{equation}
where ${\bf r}$ and ${\bf r}_i$ are points on the cap.  In what follows we will adopt  eq.~(\ref{eq_radius}) and work with spherical caps of area $\pi$: in this way our model interpolates between the Thomson problem on the unit disk~\cite{Worley06,Moore07,Amore17,Amore23}
and the Thomson problem on the sphere~\cite{Erber91,Erber97,Wales06,Wales09}.

The total electrostatic energy of the system is 
\begin{equation}
E = \sum_{i=2}^N \sum_{j=1}^{i-1} \frac{1}{r_{ij}} \ ,
\label{eq:discrete_energy}
\end{equation}
where $r_{ij}$ is the euclidean distance between any two point charges on the cap. It is well known that the number of local minima grows exponentially  with $N$ both for the case of the disk and of the sphere,  a fact that makes the search of the global minimum very challenging for large $N$. 
For the disk, an efficient strategy introduced in refs.~\cite{Amore17,Amore23} allows one to work with configurations with a definite number of border charges, $N_b$, thus reducing the computational complexity of the problem.

The approach that we have adopted here relies on an  implementation of the method of \cite{Amore17,Amore23} to deal with a spherical cap. 
We will briefly describe the procedure that we have adopted: the first observation is that the minimization of the energy, eq.~(\ref{eq:discrete_energy}),  corresponds to a {\sl constrained} optimization  problem that can be transformed to an {\sl unconstrained} one by adopting a suitable parametrization:
\begin{equation}
\begin{split}
\theta(t) &= \theta_{\rm max} \sin^2(t) \\
\phi(u) &= u
\end{split} \ .
\label{eq_parametrization}
\end{equation}

As for the Thomson problem on the disk, the border charges play an important role and typically the ground state of the system corresponds to a number of border charges with a well defined  dependence on $N$ (from the discussion  in the previous section we expect that $N_b \propto N^{2/3}$).  The algorithms devised in \cite{Amore17,Amore23} exploit this feature, by allowing one to generate configurations with a fixed number of border charges.

To enforce this condition in our scheme we split the charges into two groups: $N-N_b$ internal charges and $N_b$ border charges.  The internal charges can move along arbitrary directions but they can approach the border 
only within a distance, while border charges can only move on the border (the number of degrees of freedom in the problem is reduced to  $N_{\rm dof} =2 N-N_b$). 

The parametrization of eq.~(\ref{eq_parametrization}) can be modified into 
\begin{equation}
\begin{split}
& \left\{ 
\begin{array}{ccc}
\theta_{\rm internal}(t) &=  &\zeta \theta_{\rm max} \sin^2(t) \\
\phi_{\rm internal}(u) &= &u  \\
\end{array}\right. \\
& \left\{ 
\begin{array}{ccc}
\theta_{\rm border}(t) &=  & \theta_{\rm max}  \\
\phi_{\rm border}(u) &= &u  \\
\end{array}\right. \\
\end{split} \ , 
\label{eq_parametrization_border}
\end{equation}
where $0 < \zeta < 1$ is a parameter that constrains the internal charges to having $\theta < \theta_{\rm max}$.

Notice that $\zeta$ is not a physical parameter in the model, but rather a technical tool to speed up the search for good minima of the total energy; for this reason
it is important that the calculations are carried out using a reasonable value for $\zeta$: in fact values of $\zeta$ that are either too small or too large (too close to $1$)  may produce configurations with large energy. In particular using small  values of $\zeta$ may generate {\sl metastable} configurations, i.e. configurations that are not in equilibrium when the constraints are removed,  whereas using large values of $\zeta$, may produce configurations that could collapse to a configuration with a larger $N_b$ than expected.

As a rough estimate of $\zeta$ we can relate its value  to the average angular separation between neighboring border charges and write 
\begin{equation}
1- \zeta  \approx \frac{2\pi}{N_b} \ .
\end{equation} 
We also observe that, once a stable configuration has been found, it can be parametrized in terms of a different value of $\zeta$, if needed.

The total energy of the system is now a function of $2N-N_b$ angles and  can be minimized  using a  global optimization method such as basin--hopping: 
however it must be noted that no optimization method can guarantee that a global minimum has been reached and therefore obtaining results of good quality is typically time consuming for  large configurations.

Our strategy is a follows: 
\begin{itemize}
\item[1)] for a given value of $N$ and $\theta_{\rm max}$, we consider a range of values of $N_b$ , that likely contains the global minimum;
\item[2)] for each value of $N_b$ in this region we carry out a preliminary minimization and  use the values obtained in this way to identify (qualitatively)
the behavior of the total electrostatic energy as function of $N_b$:   if the original range of $N_b$  contains the optimal value $N_b^\star$ 
(i.e. the value corresponding to a possible global minimum), the energy will display a almost parabolic behavior close to   $N_b^\star$, superposed 
with sizable  fluctuations; usually many iterations are required to eliminate the fluctuations, thus obtaining a smoother behavior of the energy; for this reason it is not practical to carry out this process on a large set of values of $N_b$;
\item[3)] once  a smaller set of values for $N_b$ has been  identified from the previous step,  a further optimization can be carried out on this new set; depending
on the outcome of the optimization, it may be possible to further reduce the range of values of $N_b$ and repeat the optimization; 
\item[4)] at the later stages of the process one can be left with a narrow interval of values of $N_b$ ($1-3$) and 
the process can be stopped when no improvements are found in a reasonable time;
\end{itemize}

Step $2)$ above can be considerably simpler if the calculation is done sequentially, using the information for the case corresponding to $N-1$. Moreover, if one is able to obtain an accurate fit of calculated values of $N_b$ over a certain region of $N$, the fit can be used to guess relatively narrow regions of $N_b$ containing the global minimum also for larger values of $N$.

Another strategy that is particularly helpful to improve large  configurations  is the following: in a sequence of configurations spanning a region of $N_b$ (say $N_b^{min} \leq N_b \leq N_b^{max}$) it is not uncommon to observe fluctuations in the behavior of the energy as function of $N_b$  even after a good number of iterations. In this case one can select one of the configurations, that appear to have better converged  to a low energy solution and corresponding to a given $N_b$ and "transplant" it to a neighbor $N_b$ ($N_b \pm 1$). In order to generate a configuration with one more border charge, one randomly selects  an internal charge, within some distance from the border, and deposit it on the border; similarly, if one want to generate a configuration with one less border charge, one can select an arbitrary border charge and move it inside the cap. 
In any of these two cases, the ansatz configuration undergoes an energy minimization process. Overall these steps can be repeated multiple times and quite often result
in finding sizable improvements in the energy.

\section{Numerical results}
\label{sec:results} 

We have carried out a large number of numerical experiments with the purpose of finding low energy configurations for several values of $N$ and for caps of different angular width, $\theta_{\rm max}/\pi = 0.1, 0.2, \dots, 0.9$. For each of these cases, we have explored a range of values for $N_b$ (with fixed $N$),  identifying  the specific value $N_b^\star$ at which the total energy reaches the lowest minimum  (possible global minimum of the energy). Typically this process is carried out in two steps: in a first stage, one performs a preliminary exploration over a larger range of values: this phase can be safely concluded when fluctuations of the energies for the different values of $N_b$ are sufficiently small to make a reliable guess on the location of $N_b^\star$; in the next stage the search interval  is reduced and the  exploratory process  is repeated. The reduction of interval can be iterated many times until the resulting region reduces to a single value of $N_b$. 
Even when the value of $N_b^\star$ has been reliably identified, there is not assurance that the configuration found is the global minimum of the total energy: 
typically it is helpful to still carry out an number of extra trials  at $N_b=N_b^\star$.

An aspect to keep in mind is that the complexity of the problem scales exponentially with $N$:  in fact, as first observed by Erber and Hockney for the Thomson problem~\cite{Erber91, Erber97} the number of local minima grows {\sl exponentially} with $N$ (see also refs.~\cite{Calef15,Mehta16,Amore25} for more recent estimates).  As a result the exploration of the energy landscape for the Thomson problem, which to some extent is an ingredient in any endeavor of finding the global minimum, becomes  extremely hard quite rapidly. 

In general the largest configurations that we have considered have been of $N=2000$ charges, but for the special case of the hemisphere ($\theta_{\rm max} = 0.5 \pi$)  we have also studied selected configurations up to  $N=10000$ point charges. Additionally, just for the hemispherical case,  we have explored  a larger number of configurations, including all those from $N=10$ to $N=724$.
A complete catalog of the configurations found in our exploration is available at Zenodo.

\begin{figure}
\begin{center}
\includegraphics[width=\columnwidth,clip]{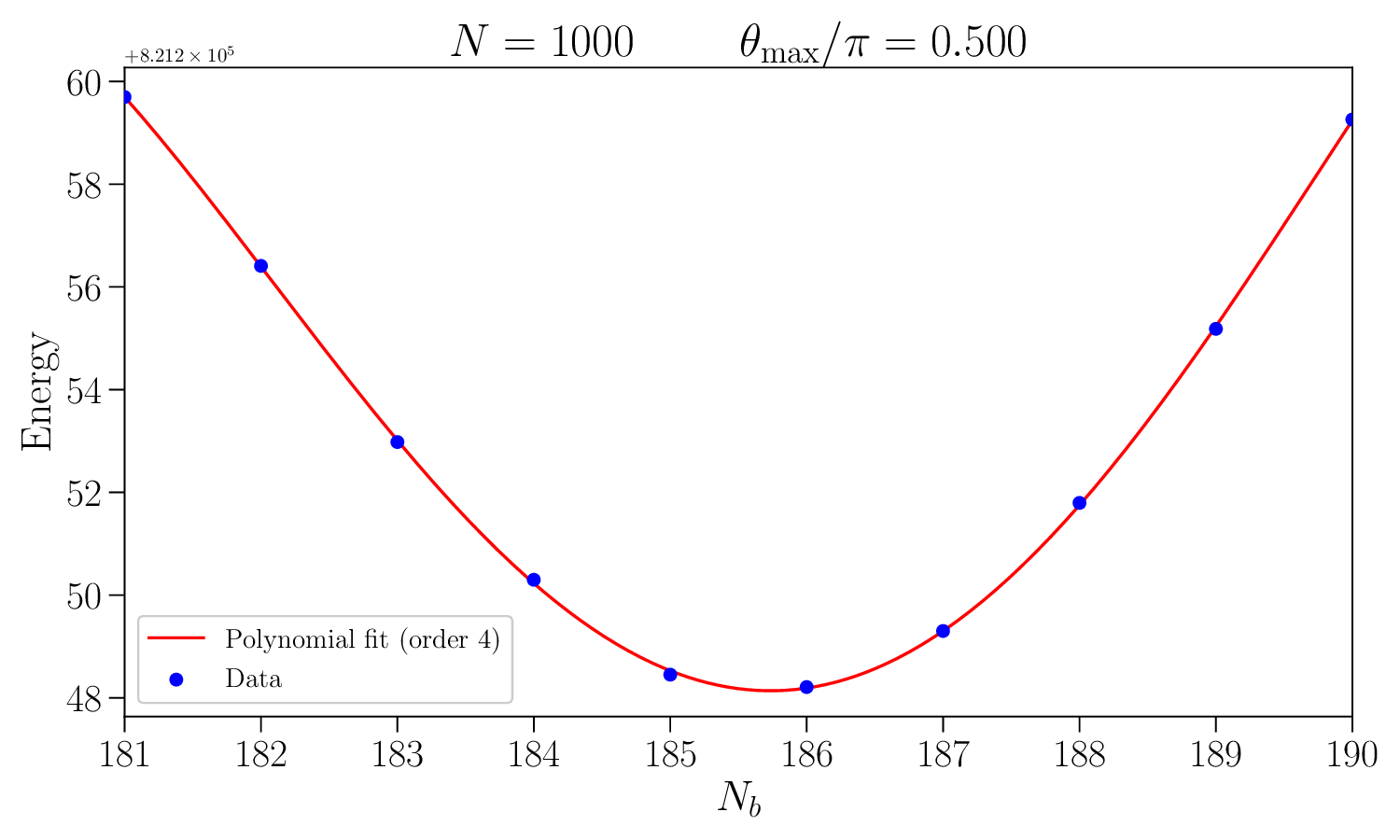}
\caption{Energy of a system of $N=1000$ charges on a spherical cap of angular width $\theta_{\rm max} = 0.5 \pi$ for different values of $N_b$.}
\label{Fig_energy_range_N_1000_thetamax_0.5}
\end{center}
\end{figure}

As an illustration of the  our results,  we discuss in detail the case of  $1000$ charges on spherical caps of angular widths  $\theta_{\rm max} =0.1 \pi$, $0.5\pi$ and $0.9\pi$.  The dependence of the total energy on the number of border charges for $\theta_{\rm max} = 0.5 \pi$  is displayed in 
 Fig.~\ref{Fig_energy_range_N_1000_thetamax_0.5} with the minimum corresponding to $N_b^\star = 186$  (the behavior for 
$\theta_{\rm max} =0.1 \pi$ and $\theta_{\rm max} =0.9 \pi$ is similar but with different minima, so we omit the plots).

\begin{figure*}[t]
    \centering
    \begin{minipage}{0.21\textwidth}\centering
        \includegraphics[width=\linewidth]{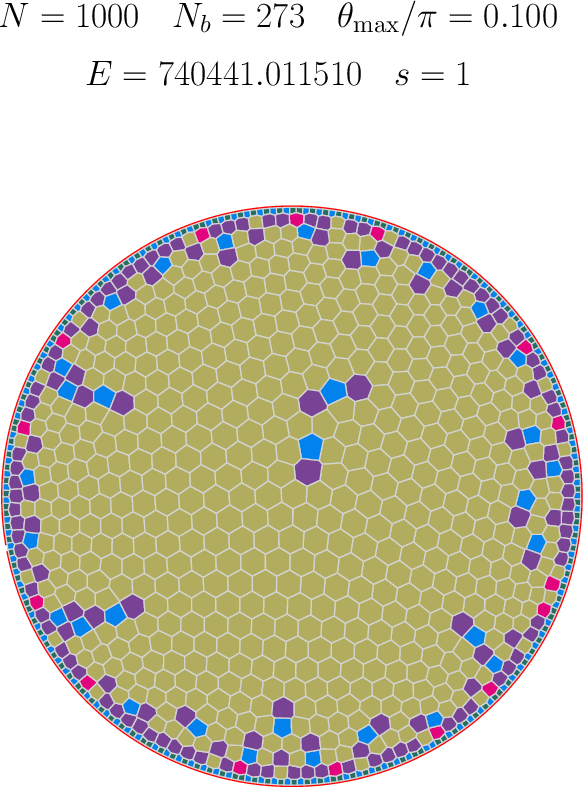}\\[-2pt]
        \scriptsize (a) $\theta_{\rm max} = 0.1\pi$, $N_b^\star = 273$
    \end{minipage}\hfill
    \begin{minipage}{0.21\textwidth}\centering
        \includegraphics[width=\linewidth]{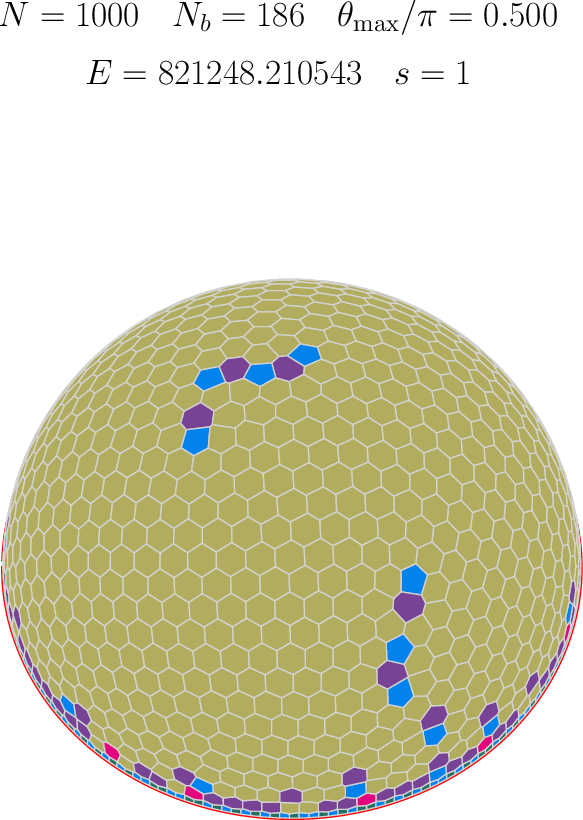}\\[-2pt]
        \scriptsize (b) $\theta_{\rm max} = 0.5\pi$, $N_b^\star = 186$
    \end{minipage}\hfill
    \begin{minipage}{0.21\textwidth}\centering
        \includegraphics[width=\linewidth]{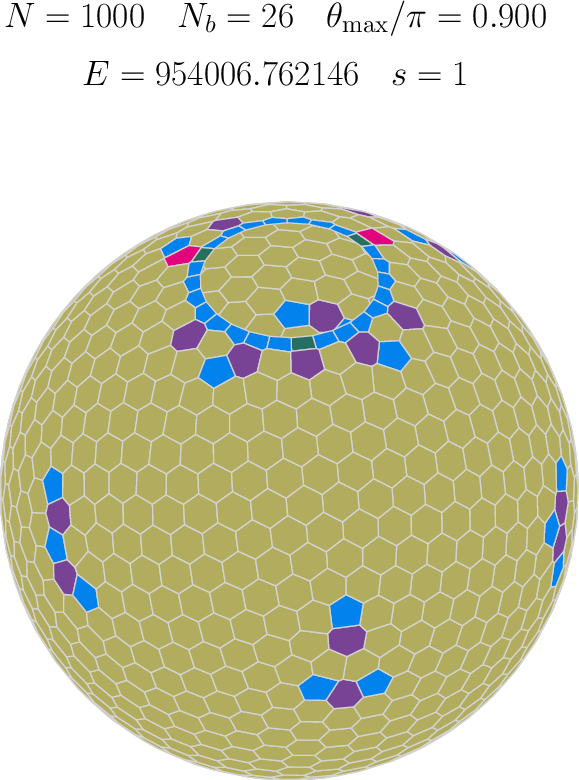}\\[-2pt]
        \scriptsize (c) $\theta_{\rm max} = 0.9\pi$, $N_b^\star = 26$
    \end{minipage}
    \caption{Configurations of $N=1000$ charges on spherical caps of increasing $\theta_{\rm max}$, with optimal $N_b^\star$.}
    \label{Fig_all_three_configs}
\end{figure*}

In Fig.~\ref{Fig_all_three_configs} we plot the Voronoi diagrams for the configurations corresponding to these three cases (the Voronoi cells are colored according to their number of sides, using the same color scheme adopted in ref.~\cite{Amore23}). Since the spherical cap is topologically equivalent to the disk (Euler's characteristic $\chi =1$) the total topological charge in the domain must add up $6$, because of Euler's theorem.

For the case $\theta_{\rm max} = 0.1 \ \pi$ we observe three chains of defects, emanating from the border in an almost radial direction (an analogous behavior is also present for  $N_b = 272$ and $N_b=274$, not shown here).  The border itself displays an almost regular alternation of pentagonal and square cells, while the next layer of cells contains predominantly heptagonal cells, with a minority of octagonal and hexagonal shapes: this behavior is very similar to the one observed in the flat disk.
The size of the border cells is considerably smaller than the size of the internal cells, as expected from the density of the continuum problem, that has an integrable singularity at the border. 

The best configuration found for $\theta_{\rm max} = 0.5 \ \pi$ (part (b) of the figure) displays a large isolated positively charged scar close to the center and 
a sequence of defects emanating from the border and almost perpendicularly to it. The border cells display a similar alternation of pentagons and squares, with a layer of predominantly heptagonal cells next to it.
 
Finally, for the best configuration for $\theta_{\rm max}  = 0.9 \ \pi$  we observe that  the border cells are predominantly pentagonal,  resulting in a topological border charge is small and positive (the next layer of cells is now formed principally by hexagonal (neutral) cells, with  a small number of heptagons and octagons).   The bulk, in this case, carries a larger positive topological charge, as required from Euler's theorem, with a distribution of defects that starts to resemble that on the sphere.

In what follows we focus on analyzing the great amount of data that we have produced: this analysis has been carried out for all the spherical caps considered in this paper, but we will only show plots  for the hemisphere, for reason of space.

\begin{figure}
\begin{center}
\includegraphics[width=\columnwidth,clip]{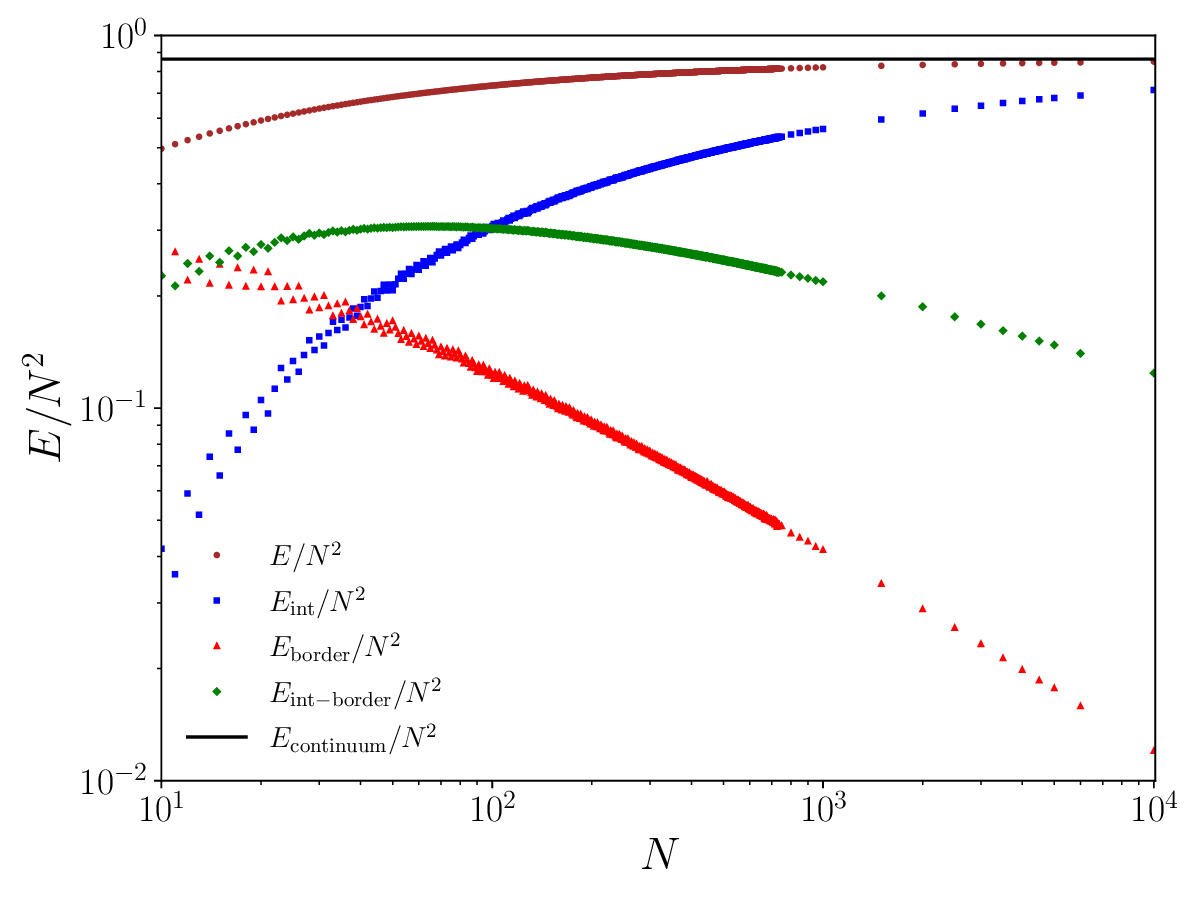} 
\caption{Electrostatic energy  (divided by $N^2$) as a function of $N$ for $\theta_{\rm max} = 0.5 \pi$: the different sets in the plot correspond to the full discrete component, the component relative to the border charges, to the internal charges and to the interaction between border and internal charges. The horizontal line corresponds to the continuum result. }
\label{Fig_energies_05}
\end{center}
\end{figure}

In Fig.~\ref{Fig_energies_05} we plot the total electrostatic energy of the spherical cap  with $\theta_{\rm max} = 0.5\pi$, divided by $N^2$, for the different values of $N$
for which we have found numerical results; for the larger $N$, the curve can be seen to approach the result for the continuum  quite closely (horizontal line). The interaction energies of the border charges, of the internal charges and of internal and border charges are also displayed in the figure  and show that for 
$N \approx 10^2$ the energy of internal charges starts to dominate.

\begin{figure}
\begin{center}
\includegraphics[width=8cm]{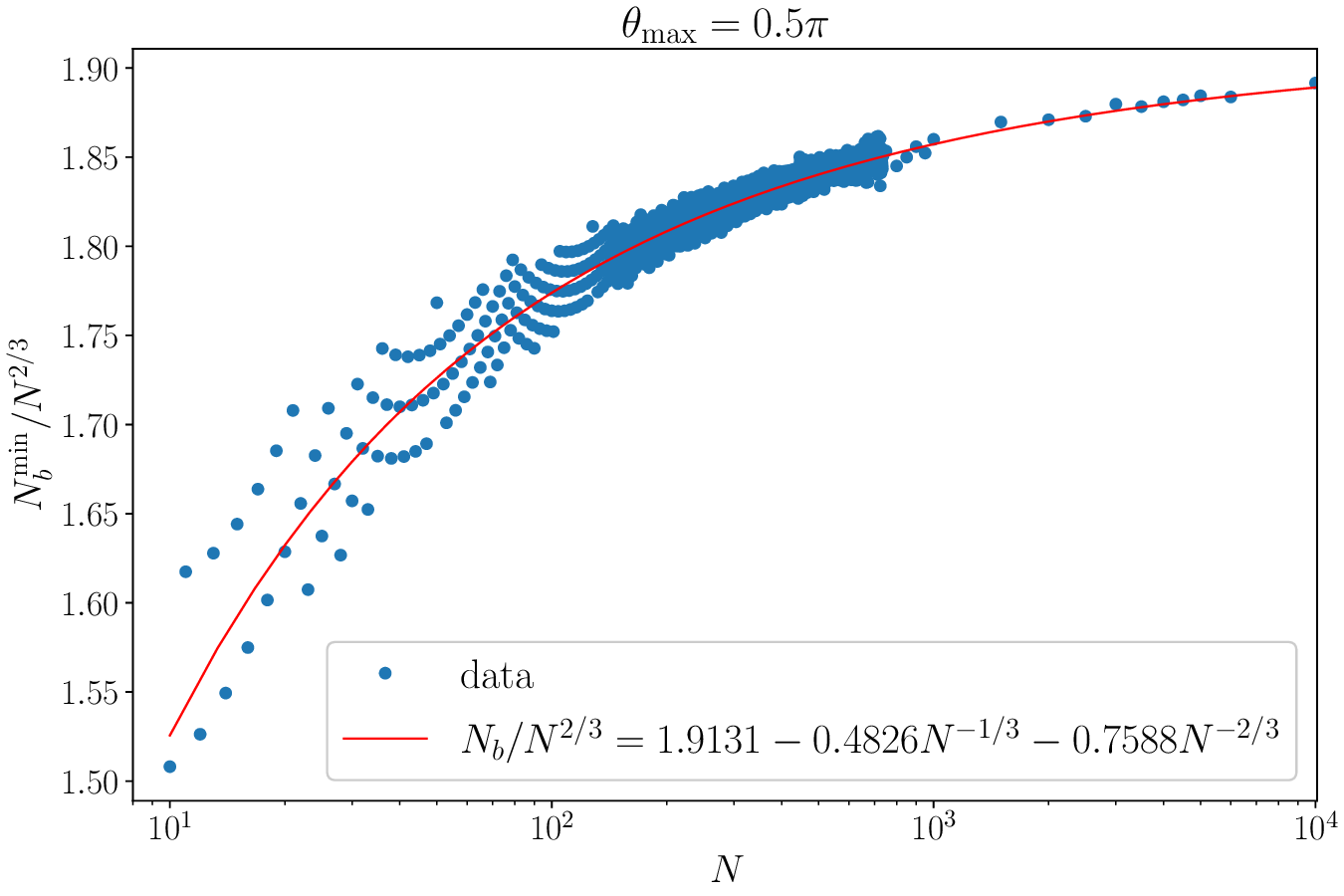} 
\caption{$N_b/N^{3/2}$ as a function of $N$ for $\theta_{\rm max} = 0.5 \pi$.}
\label{Fig_Nb_2_vs_N}
\end{center}
\end{figure}

In Fig.~\ref{Fig_Nb_2_vs_N} we plot $N_b/N^{3/2}$ as a function of $N$ for $\theta_{\rm max} = 0.5 \pi$; 
the data are well described by the fit of the form $N_b = \lambda N^{2/3} + \lambda_1 N^{1/3} +\lambda_2$
(continuous line in the plot)~\footnote{Notice however that the values of $N_b$ for $N=5000$ and $N=10000$ are not inputs in the plot, but obtained from the fit itself.}.
\begin{table}
\begin{center}
\begin{tabular}{|c|c||}
  \hline
  $\theta_{\rm max}/\pi$ & $N_b^{(fit)}$ \\
  \hline
$0.1$ & $2.8212 N^{2/3} - 0.7375 N^{1/3}-1.5869$ \\
$0.2$ & $2.7103 N^{2/3} - 0.8434 N^{1/3}-0.9592$ \\
$0.3$ & $2.4920 N^{2/3} - 0.4302 N^{1/3}-1.9388$ \\
$0.4$ & $2.2569 N^{2/3} - 0.7483 N^{1/3}-0.5131$ \\
$0.5$ & $1.9131 N^{2/3} - 0.4826 N^{1/3}-0.7588$ \\
$0.6$ & $1.5200 N^{2/3} - 0.2759 N^{1/3}-0.7674$ \\
$0.7$ & $1.0955 N^{2/3} - 0.137 N^{1/3}-0.7374$ \\
$0.8$ & $0.6906 N^{2/3} - 0.4589 N^{1/3}+0.7850$ \\
$0.9$ & $0.2330 N^{2/3} - 0.4649 N^{1/3}-2.0168$ \\
  \hline
\end{tabular}
  \caption{Parameters of the fit  of $N_b$ for different values of $\theta_{\rm max}$.} 
  \label{tab:table_lambda}  
 \end{center}
  \end{table}

The fits for the different values of $\theta_{\rm max}$ are reported in Table \ref{tab:table_lambda}. We expect that the fits are less precise as $\theta_{\rm max} \rightarrow \pi$, because in this limit $N_b$ is small for the values of $N$ that we are able to consider in this paper. 

\begin{figure}
\begin{center}
\includegraphics[width=8cm]{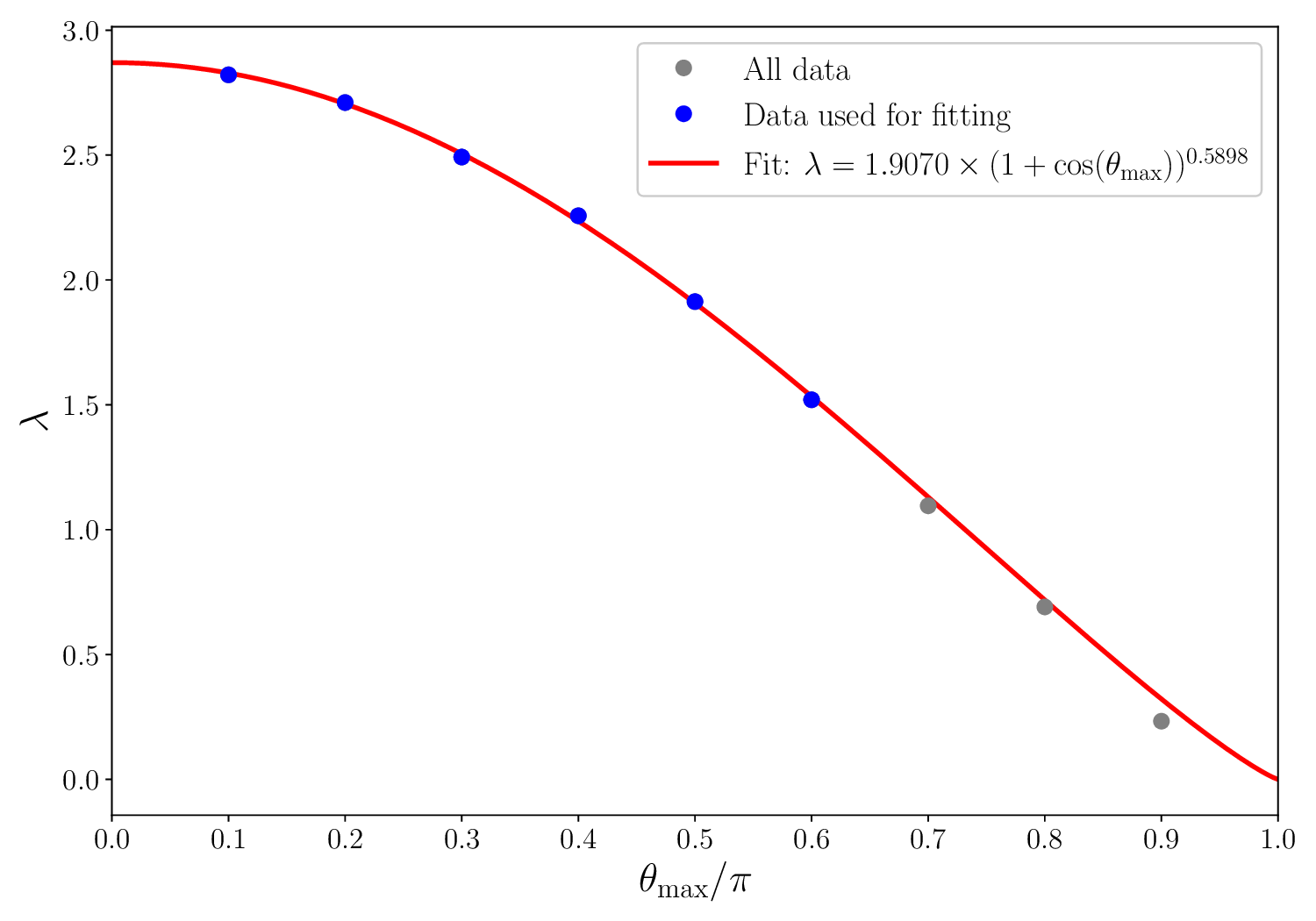} 
\bigskip
\bigskip
\caption{$\lambda$ versus $\theta_{\rm max}$.}
\label{Fig_lambda_vs_thetamax}
\end{center}
\end{figure}

The leading coefficient in these fits, $\lambda$, is plotted in  Fig.~\ref{Fig_lambda_vs_thetamax}; the continuous curve in the plot is the 
fit
\begin{equation}
\lambda = 1.907 \left( 1 + \cos\theta_{\rm max}\right)^{0.5898} \nonumber \ ,
\end{equation}
obtained using the data from $0.1 \leq  \theta_{\rm max}/\pi \leq 0.6$.

\begin{table}
\begin{center}
\begin{tabular}{|c|c|c|c||c|}
  \hline
  $\theta_{\rm max}/\pi$ & $\kappa_2$ & $b$ & $c$ & $\kappa_2^{\rm LDA}$ \\
  \hline
$0.1$ & -1.5546 & -0.4397 &   0.1920 & -1.5563 \\
$0.2$ & -1.5317 & -0.4155 &   0.2102& -1.5334 \\
$0.3$ & -1.4944 & -0.2908 &   0.330 & -1.4956 \\
$0.4$ & -1.4433 & -0.0682 &   0.5442  &  -1.4442\\
$0.5$ & -1.3801 & 0.0053 &   0.5739 & -1.3811\\
$0.6$ & -1.3054 & -1.0173 &   - 0.573 &  -1.3100\\
$0.7$ & -1.2362 & 0.2987 &  0.7481  & -1.2370 \\
$0.8$ & -1.1705 & 1.2057 &  1.5628  & -1.1711\\
$0.9$ & -1.123 & 2.8189 &  3.0262 & -1.1237\\
  \hline
\end{tabular}
  \caption{Parameters of the fit  of the energy for different values of $\theta_{\rm max}$.} 
  \label{tab:table_fit_energy}  
\end{center}
  \end{table}

\begin{figure}
\begin{center}
\includegraphics[width=\columnwidth,clip]{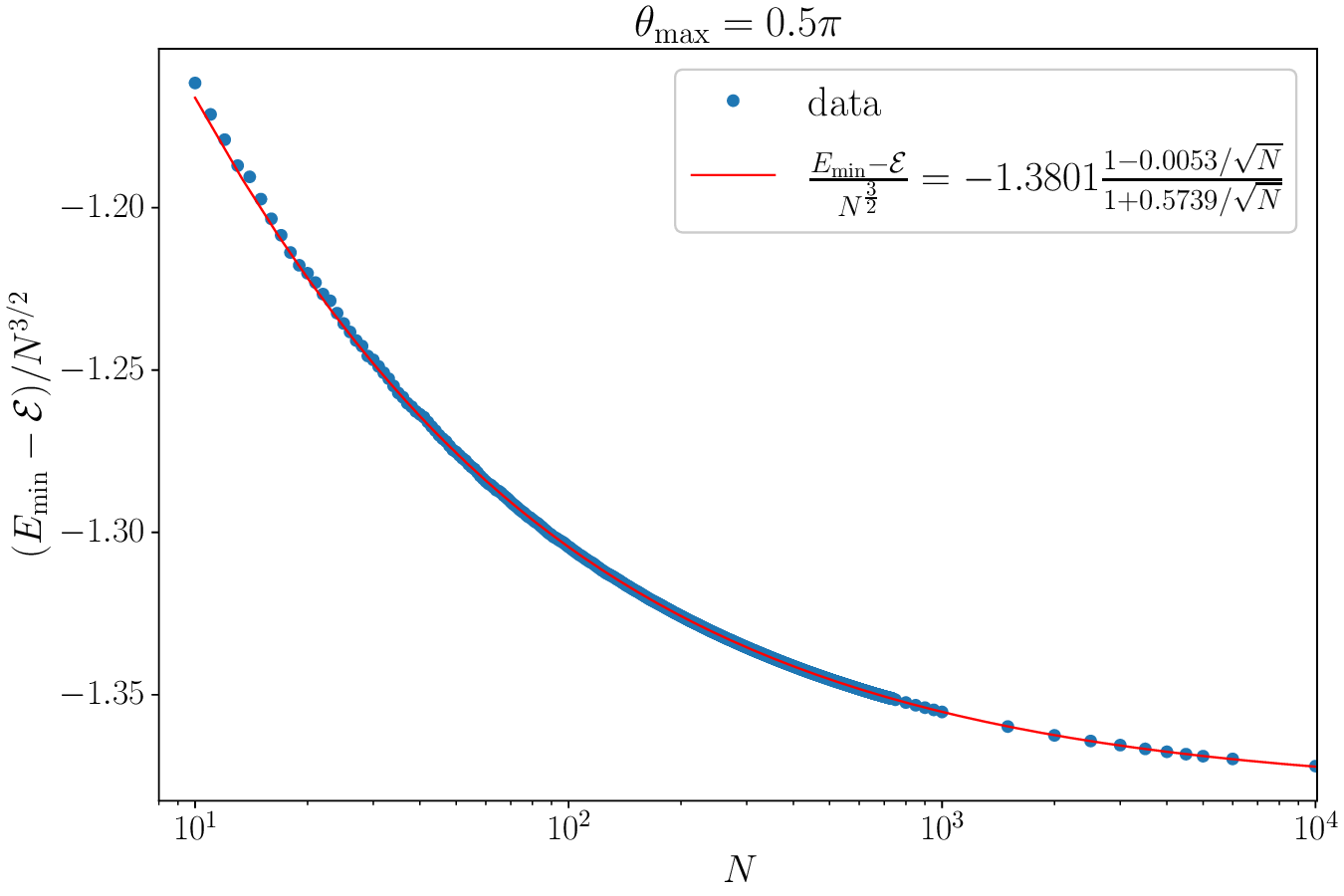} 
\caption{$\left( E_{\rm min}- \mathcal{E}\right)/N^{3/2}$ as a function of $N$.}
\label{Fig_delta_energy_0.5}
\end{center}
\end{figure}

\begin{figure}
\begin{center}
\includegraphics[width=\columnwidth,clip]{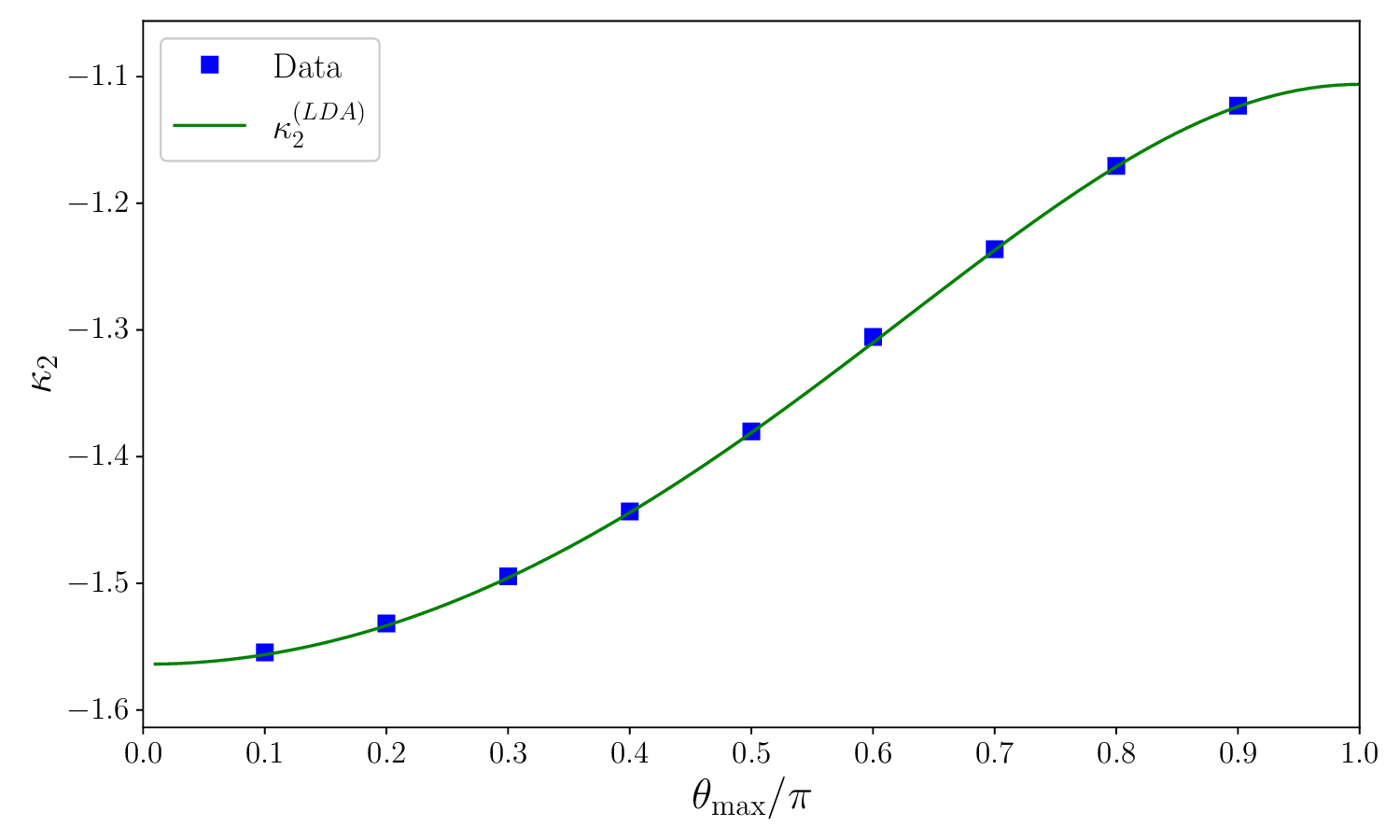} 
\caption{$\kappa_2$ versus $\theta_{\rm max}$. The solid curve is the result obtained using LDA  with $\beta_\Delta = 3.921034$.}
\label{Fig_kappa2_vs_thetamax}
\end{center}
\end{figure}

As expected, the energies of the discrete configurations  grow as $N^2$ for large $N$ and the numerical results are well described by the behavior
\begin{equation}
\begin{split}
E_{smooth}(\theta_{\rm max}, N) &= \mathcal{E}(\theta_{\rm max}, N)  \\
&+  \kappa_2(\theta_{\rm max})   N^{3/2}  \frac{(1 + b(\theta_{\rm max})/\sqrt{N})}{(1 + c(\theta_{\rm max})/\sqrt{N})} \ ,
\end{split}
\end{equation}
where the coefficients are obtained from  a least square fit and $\mathcal{E}(\theta_{\rm max}, N)$ is the energy of a continuous distribution of charge reported in
eq.~(\ref{eq:Econtinuum})

For instance, for $\theta_{\rm max}=\pi/2$ we have found:
\begin{equation}
\begin{split}
E_{\rm smooth}(\theta_{\rm max}, N) &\approx  \mathcal{E}(\theta_{\rm max}, N)  + \kappa_2  N^{3/2} +\kappa_2  (b-c)  N \\
&+ \kappa_2  \left(c^2-b c\right) \sqrt{N}  +\kappa_2  \left(b c^2-c^3\right) + \dots \\
&\approx \frac{\sqrt{2}  \pi  N^2}{2+\pi }-1.3801 N^{3/2} \\
&+ 0.784725   N-0.450354 \sqrt{N} +\dots
\end{split}
\end{equation}

The dominant term in this expansion is the {\sl correlation energy}:  Mughal and Moore~\cite{Moore07} have estimated it
using the local density approximation (LDA) for the case of the flat disk; their formula can be adapted to the spherical cap as
\begin{equation}
E^{(LDA)}_{\rm corr} = - \frac{\beta}{2 R(\theta_{\rm max})} \int_0^{\theta_{\rm max}} \left[ \sigma(\theta) \right]^{3/2}  2\pi \sin(\theta) d\theta  \ ,
\end{equation}
where $\beta$ depends on the geometric properties of the lattice ($\beta_\Delta=3.921034$ for a triangular lattice~\cite{Bonsall77, Moore07}).

Mughal and Moore had found an excellent agreement between the value of $\kappa_2$ estimated from their numerical results and the one obtained from LDA using 
$\beta_\Delta$: as we can see from Fig.~\ref{Fig_kappa2_vs_thetamax} the LDA reproduces quite well the dependence of $\kappa_2$ over the angular width $\theta_{\rm max }$ (solid curve in the plot).

\begin{figure}
\begin{center}
\includegraphics[width=8cm]{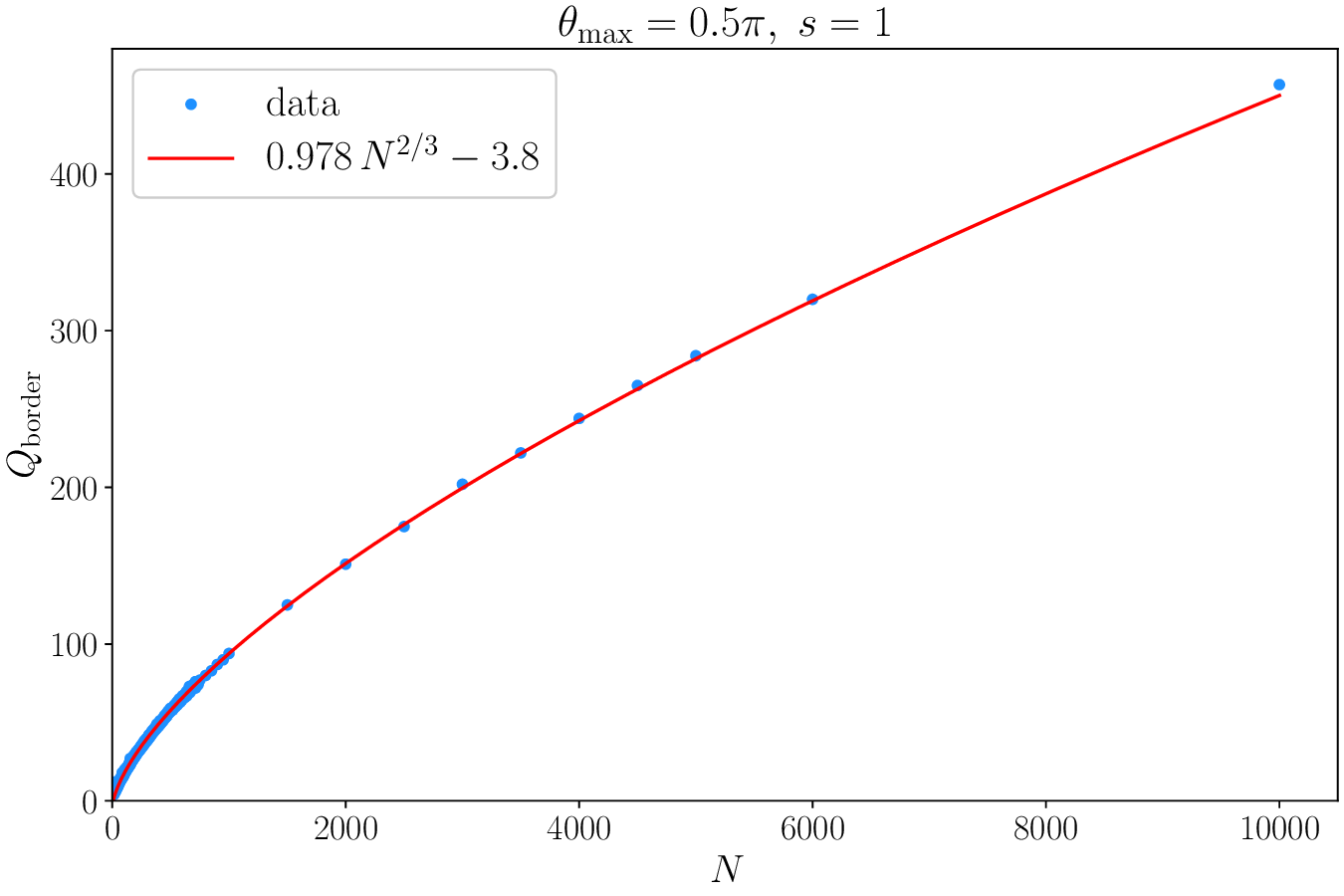}
\caption{Border topological charge versus $N$  on a hemisphere ($\theta_{\rm max}=0.5 \pi$).}
\label{Fig_Qb_vs_N_05}
\end{center}
\end{figure}

\begin{figure}
\begin{center}
\includegraphics[width=8cm]{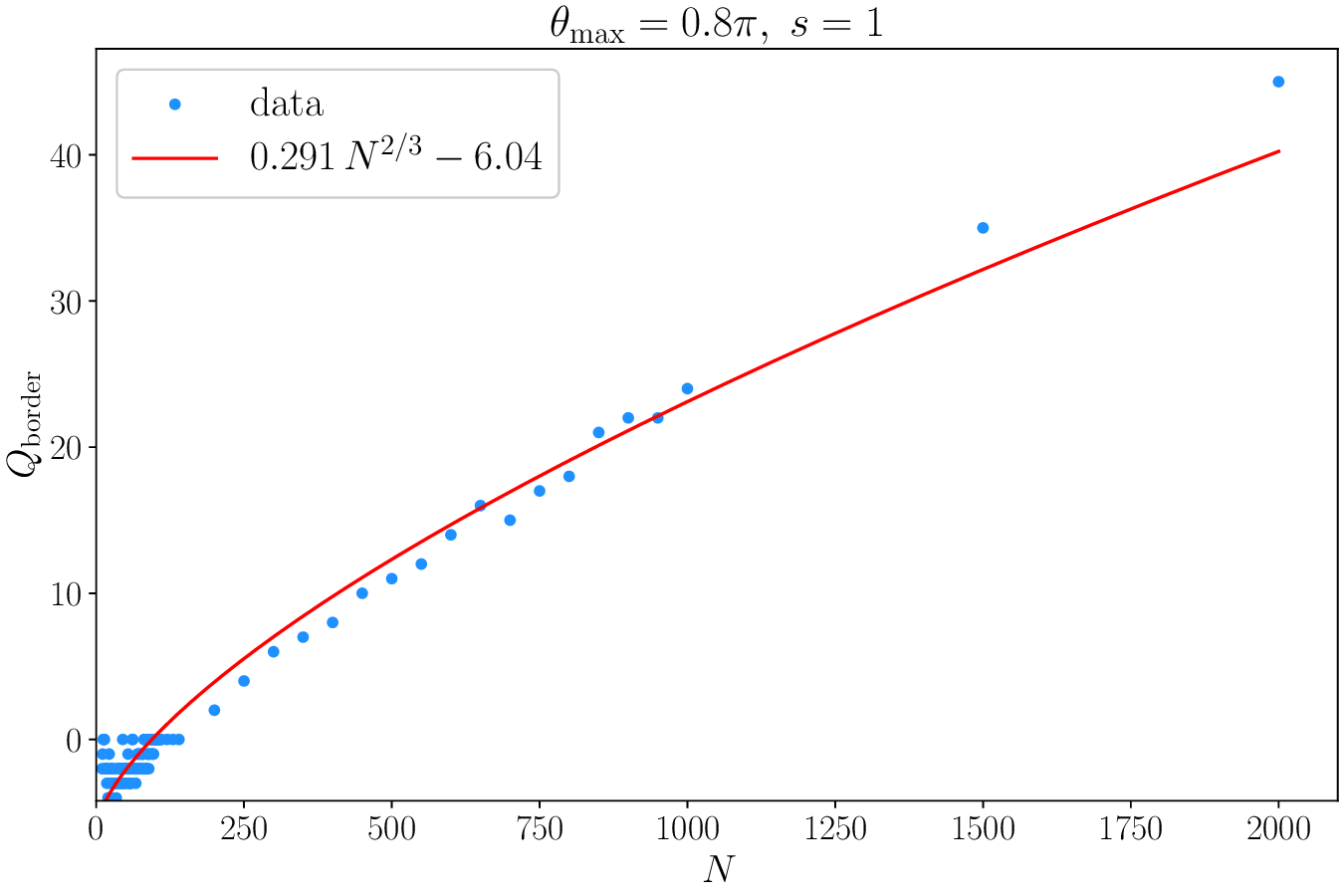}
\caption{Border topological charge versus $N$  on a hemisphere ($\theta_{\rm max}=0.8 \pi$).}
\label{Fig_Qb_vs_N_08}
\end{center}
\end{figure}

In Fig.~\ref{Fig_Qb_vs_N_05} and \ref{Fig_Qb_vs_N_08} we plot the border topological charge for $\theta_{\rm max} = 0.5 \pi$ and $0.8 \pi$, as a function of $N$. 
The fits of the kind $a N^{2/3}+ b$ (solid lines in the plots) describe rather well the dependence of $Q_{\rm border}$ on $N$, particularly for the cap of smaller width. 
There is a simple explanation to this behavior: from the Voronoi diagrams of the configurations with $\theta_{\rm max } = 0.5$  we appreciate that the border 
is covered by pentagonal and square cells, in an almost perfect alternation. This means that there are roughly $N_b^{\rm (4)} \approx N_b/2$ square cells on the border and the border topological charge is  $Q_{\rm border} \approx N_b/2$, which is responsible for the $N^{2/3}$ growth observed in the plot.

The case for $\theta_{\rm max} = 0.8\pi$ is more complex: in this case we observe that for $N \leq 140$ the border  topological charge is actually null or negative (for instance, at $N=140$ the border cells are just pentagons).   The behavior at larger $N$ can still be (qualitatively) described by the fit. Notice however that the numerical experiments in this case have been carried out on a smaller range of values of $N$.

\begin{figure}
\begin{center}
\includegraphics[width=8cm]{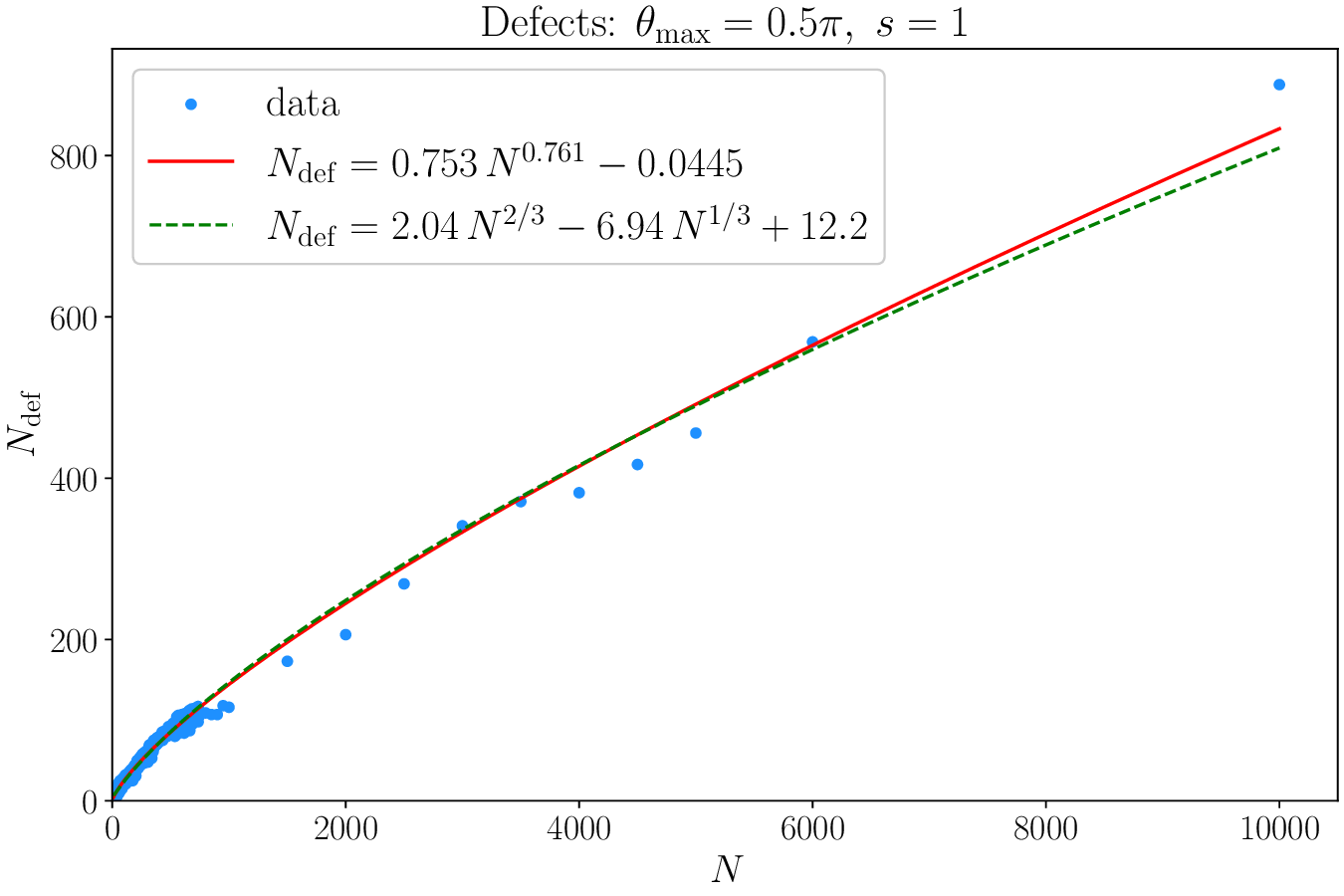}
\caption{Number of internal defects versus $N$  on a hemisphere ($\theta_{\rm max}=0.5 \pi$).}
\label{Fig_Ndef_vs_N_05}
\end{center}
\end{figure}

In Fig.~\ref{Fig_Ndef_vs_N_05} we plot the  number of internal defects for configurations on a hemisphere as function of the number of border charges, $N_b$. Euler theorem requires that the topological charge of the cap be always equal to $6$~\footnote{Although this result is general, one should take into account that vertices of order larger than $3$ carry a topological charge; for a discussion of this point see \cite{Amore23b}. }, but this does not constrain the number of defects itself: large scars with an equal number of pentagonal and heptagonal cells, for instance, do not alter the total topological charge and therefore are allowed  (as a matter of fact, for the Thomson problem on the sphere it has been observed that the nature  of defects appearing in the configurations is changing as $N$ grows~\cite{Wales06,Wales09}). 

The lines in the plot correspond to two different three--parameter fits, of comparable quality (the exponent $2/3$ in the second plot is justified by our previous analysis on the behavior of $Q_b$). An exponent larger than $2/3$ may signal the tendency for large $N$ to have an excess of defects carrying a null topological charge.

\begin{figure}[t]
    \centering
    \includegraphics[width=0.4\columnwidth,clip]{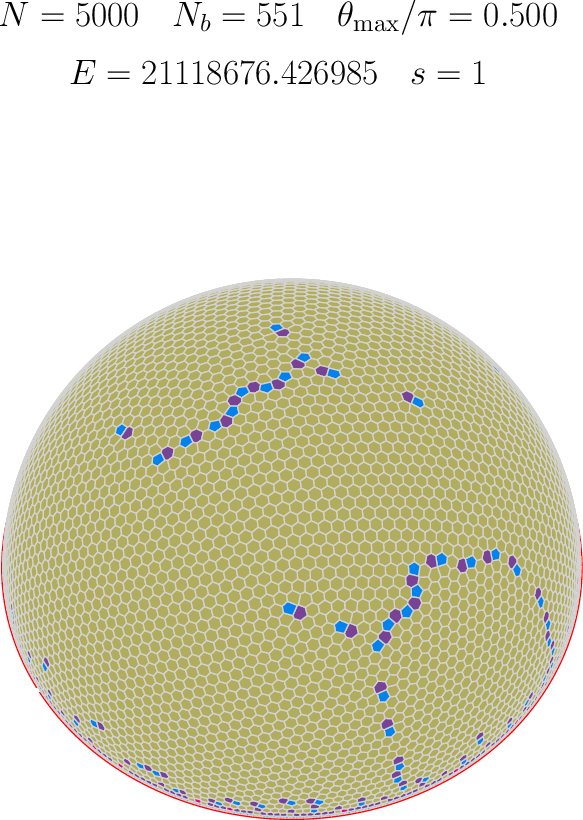}
    \caption{Low energy configuration of $N=5000$ point charges on a hemisphere ($\theta_{\rm max}=0.5 \pi$).}
    \label{Fig_5000_thetamax_05}
\end{figure}

\begin{figure}[t]
    \centering
    \includegraphics[width=0.7\columnwidth,clip]{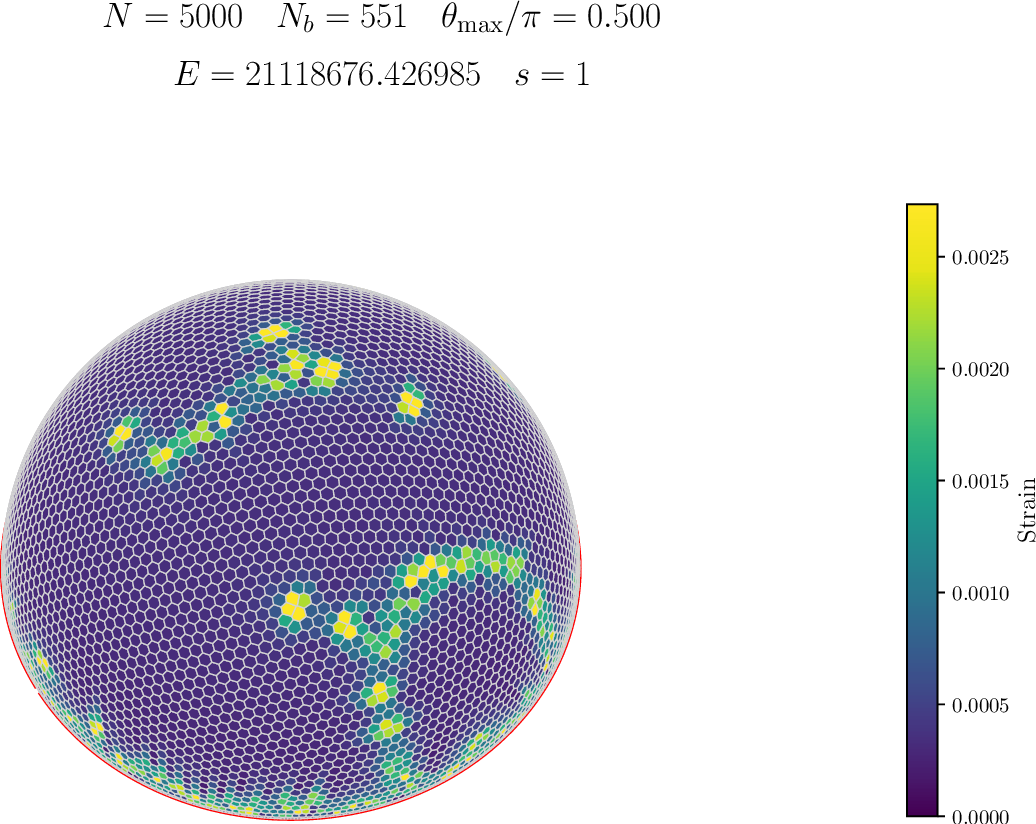}
    \caption{Low energy configuration of $N=5000$ point charges on a hemisphere ($\theta_{\rm max}=0.5\pi$).}
    \label{Fig_5000_thetamax_05_strain}
\end{figure}

In Fig.~\ref{Fig_5000_thetamax_05} we plot the lowest energy configuration that we have found for $5000$ point charges on the hemisphere, corresponding to $N_b =551$.  This configuration displays large sequences of defects, in a sea of hexagonal cells. Fig.~\ref{Fig_5000_thetamax_05_strain} displays the same configuration, but in this case the Voronoi cells are colored according to their strain: we define the strain as the distance between the center of mass of the Voronoi cell and the actual position of  the point charge.
Cells that correspond to a regular spherical polygon, for instance, would have a vanishing strain: the presence of border and curvature make it impossible to produce configurations
without strain and in fact we observe that regions with larger strain correspond roughly to the regions with defects (and immediate surroundings).

\begin{figure}
\begin{center}
\includegraphics[width=\columnwidth,clip]{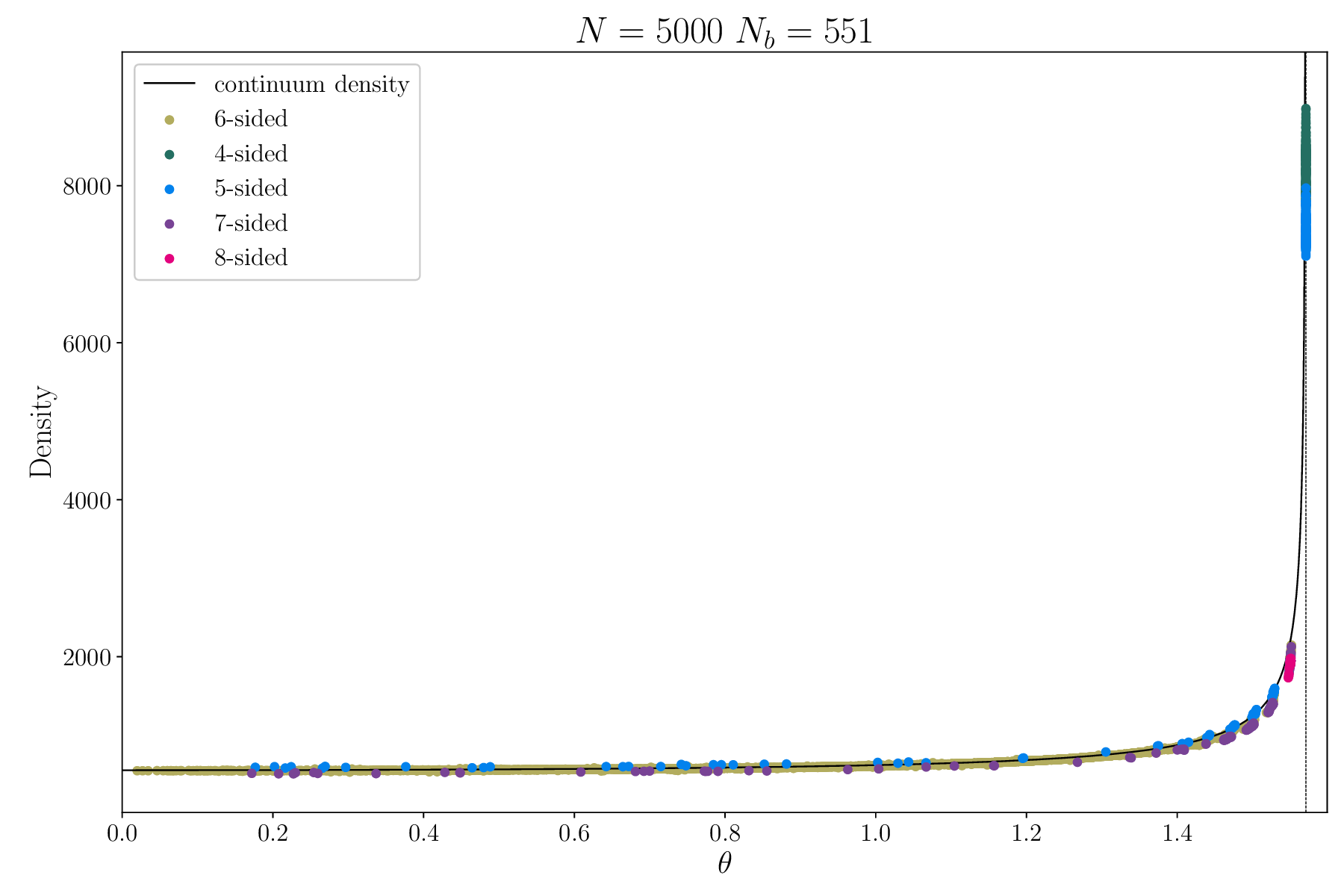}
\caption{Charge density for the configuration of Fig.~\ref{Fig_5000_thetamax_05}.}
\label{Fig_density_5000_05}
\end{center}
\end{figure}

The density corresponding to this configuration is displayed in Fig.~\ref{Fig_density_5000_05} and compared with the result for the
continuum distribution (solid line in the plot). The discrete density at each Voronoi cell is obtained as $1/a_i$ where $a_i$ is the area of the cell.

\begin{figure}
\begin{center}
\includegraphics[width=\columnwidth,clip]{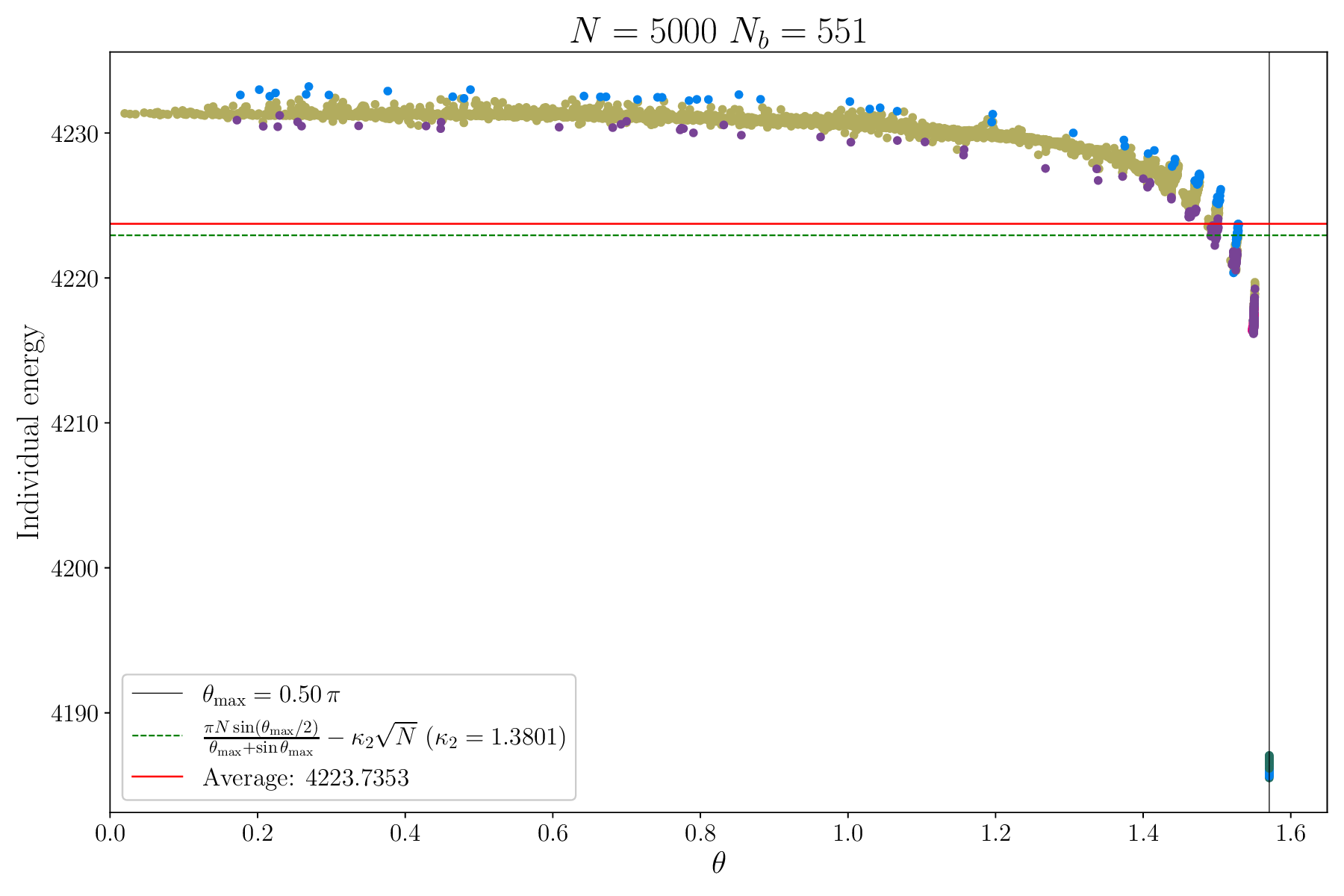}
\caption{Individual energies for a low energy configuration of $N=5000$ point charges on a hemisphere.}
\label{Fig_individual_full_05}
\end{center}
\end{figure}

\begin{figure}
\begin{center}
\includegraphics[width=\columnwidth,clip]{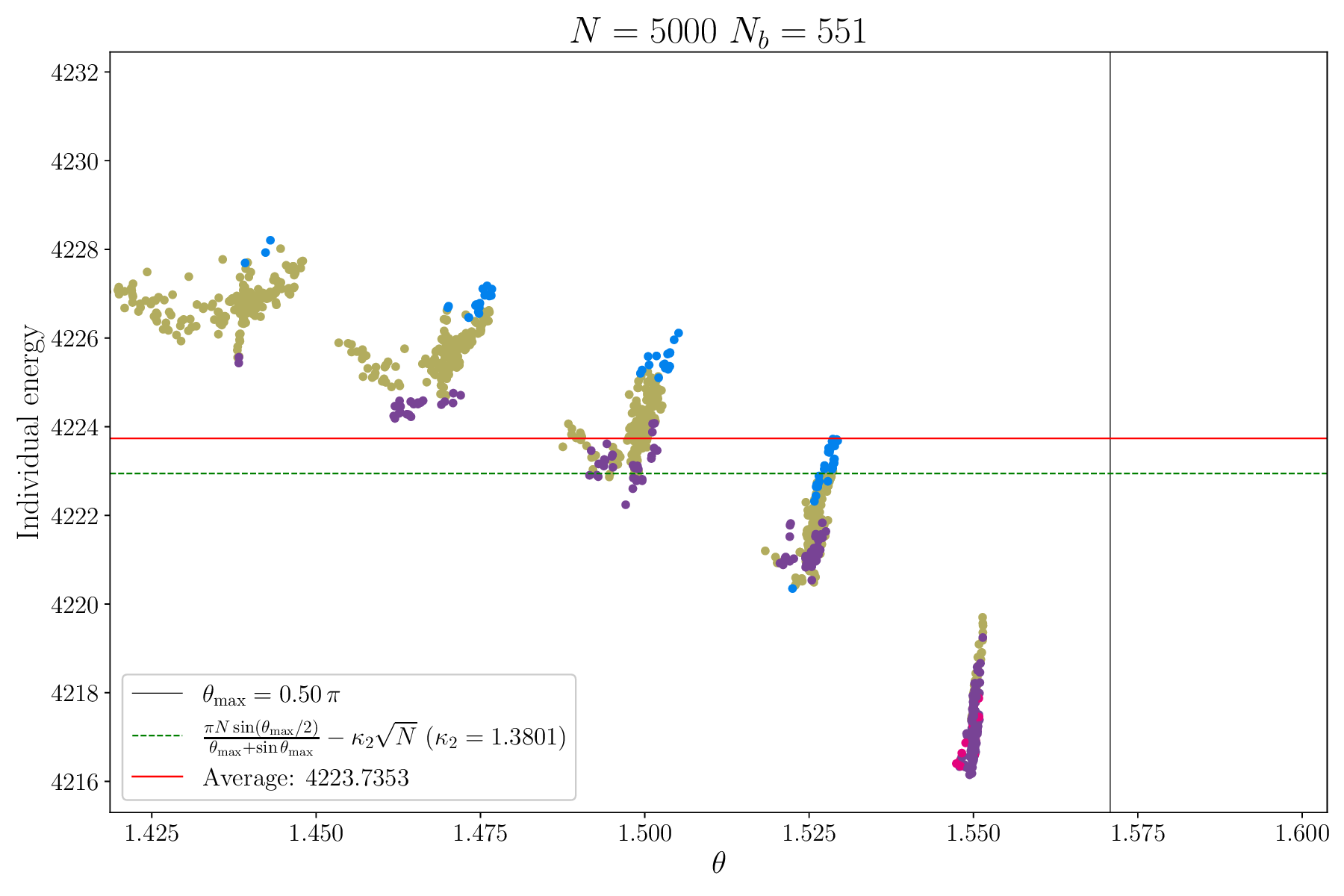}
\caption{Individual energies for a low energy configuration of $N=5000$ point charges on a hemisphere (detail).}
\label{Fig_individual_detail_05}
\end{center}
\end{figure}

In Fig.~\ref{Fig_individual_full_05} we have plotted the individual energies for the point charges in the configurations  of  Fig.~\ref{Fig_5000_thetamax_05}, coloring 
them according to the number of sides of the corresponding Voronoi cells.  The energies decay as one departs from the north pole, with the lowest values being reached at the border of the cap: there is  a sizable energy gap separating the border points with the immediate neighbors.
It can also be appreciated that, in general,  pentagonal cells have a higher energy than hexagonal and heptagonal cells
and that, close to the border, the configuration displays a clear shell structure, with four well separated layers  (see Fig.~\ref{Fig_individual_detail_05}), which contain most of the internal defects: the first layer, closest to the border, also contains octagonal cells.

\begin{figure}
\begin{center}
\includegraphics[width=\columnwidth,clip]{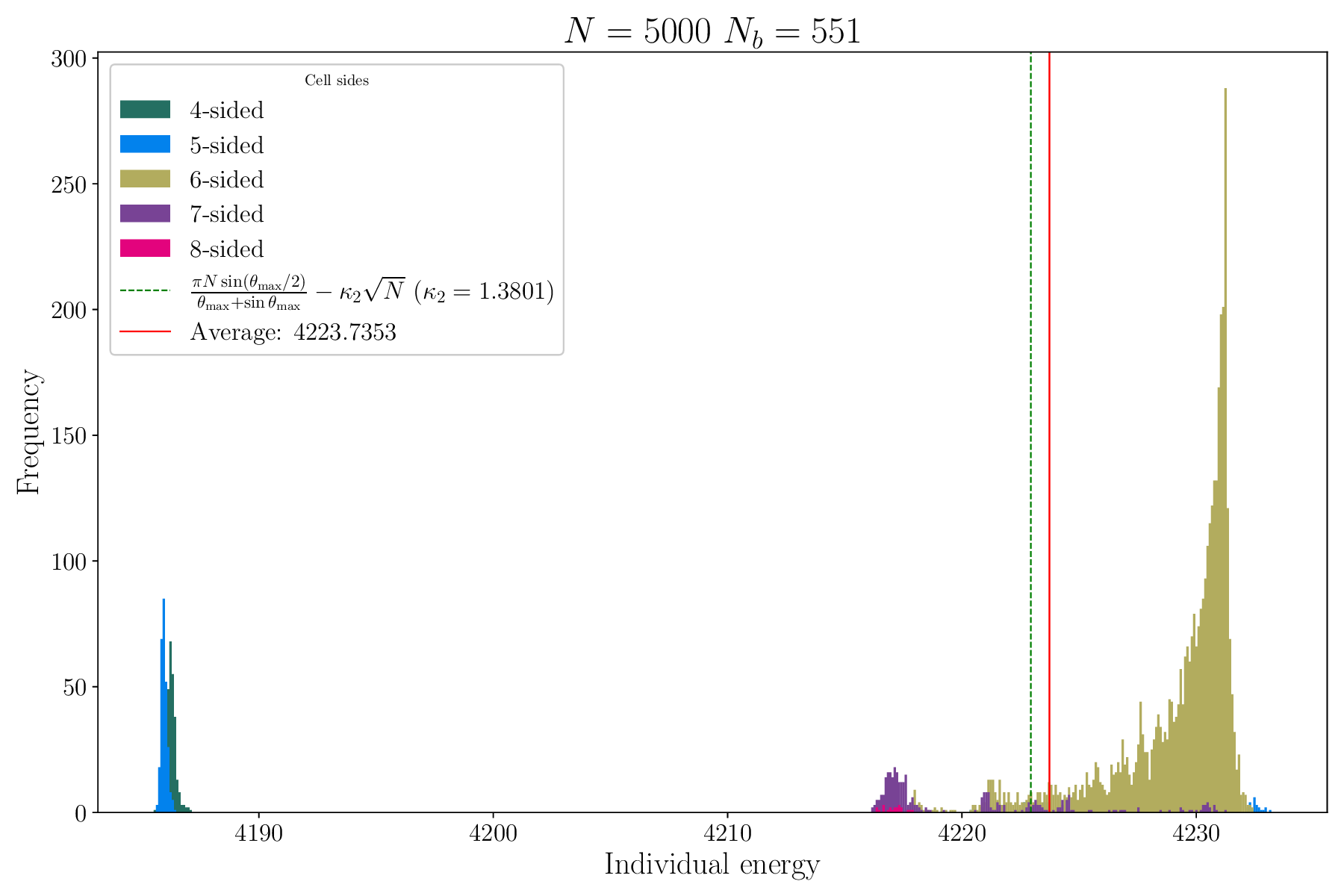}
\caption{Histogram of individual energies for a low energy configuration of $N=5000$ point charges on a hemisphere.}
\label{Fig_individual_histogram_05}
\end{center}
\end{figure}

\begin{figure}
\begin{center}
\includegraphics[width=\columnwidth,clip]{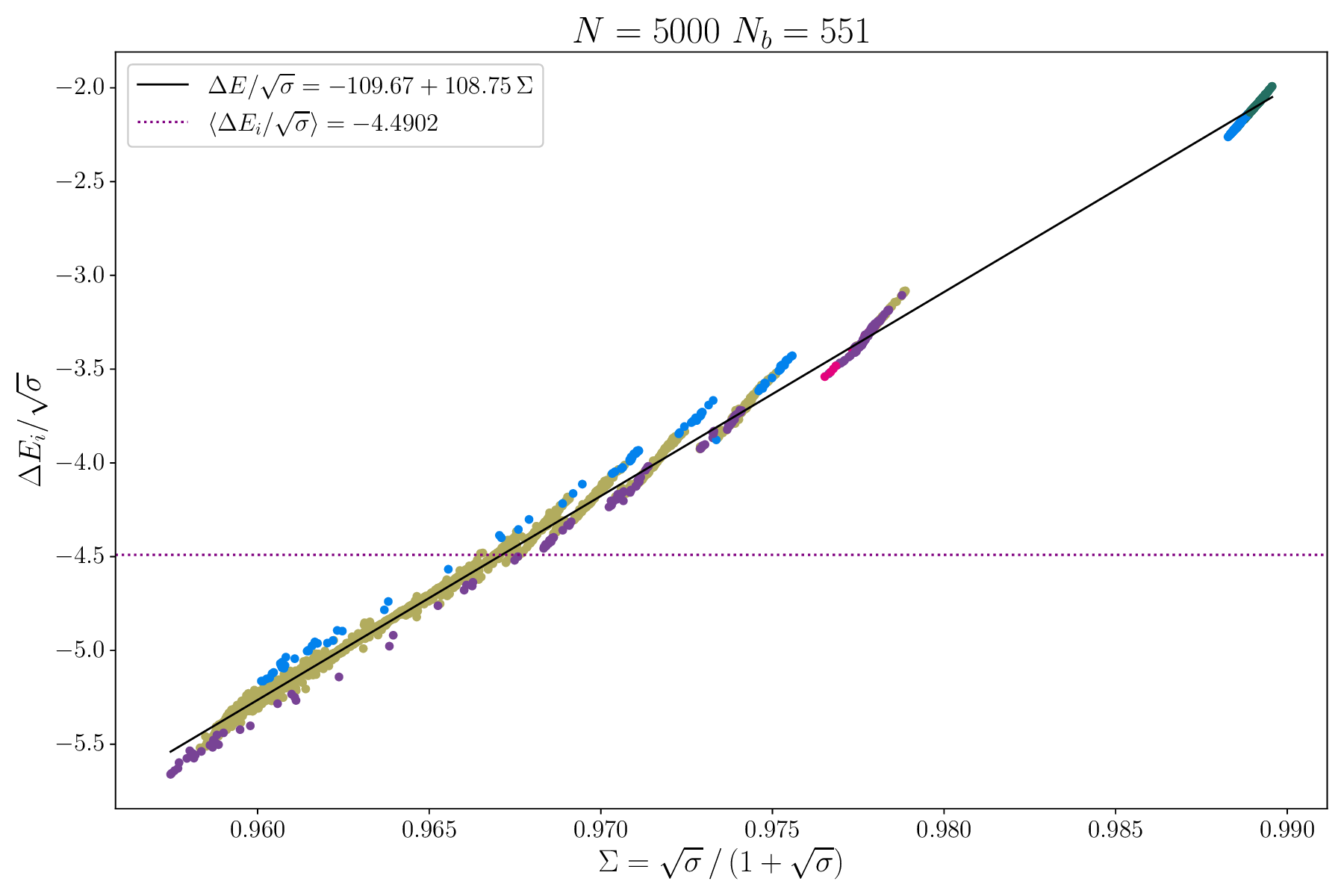}
\caption{Individual energies for a low energy configuration of $N=5000$ point charges on a hemisphere (detail).}
\label{Fig_individual_deltaE_05}
\end{center}
\end{figure}

Our figure \ref{Fig_individual_deltaE_05} for the configuration of $5000$ points of Fig.~\ref{Fig_5000_thetamax_05}  is analogous to Fig.5 of \cite{Moore07} for the flat disk, with the difference that the quantity $\Delta E_i /\sqrt{\sigma} = (E_i -\mu)/\sqrt{\sigma}$ is plotted as a function of the variable $\Sigma \equiv \sqrt{\sigma}/(1+\sqrt{\sigma})$ (at the border of the cap $\Sigma = 1$). Here $\sigma$ is the continuum density for $\theta_{\rm max } = \pi/2$. The solid line in the plot is the linear fit $\Delta E/\sqrt{\sigma}  = -109.67 + 108.75 \ \Sigma$, which describes the data rather well.

\begin{figure}
\begin{center}
\bigskip\bigskip\bigskip
\includegraphics[width=\columnwidth,clip]{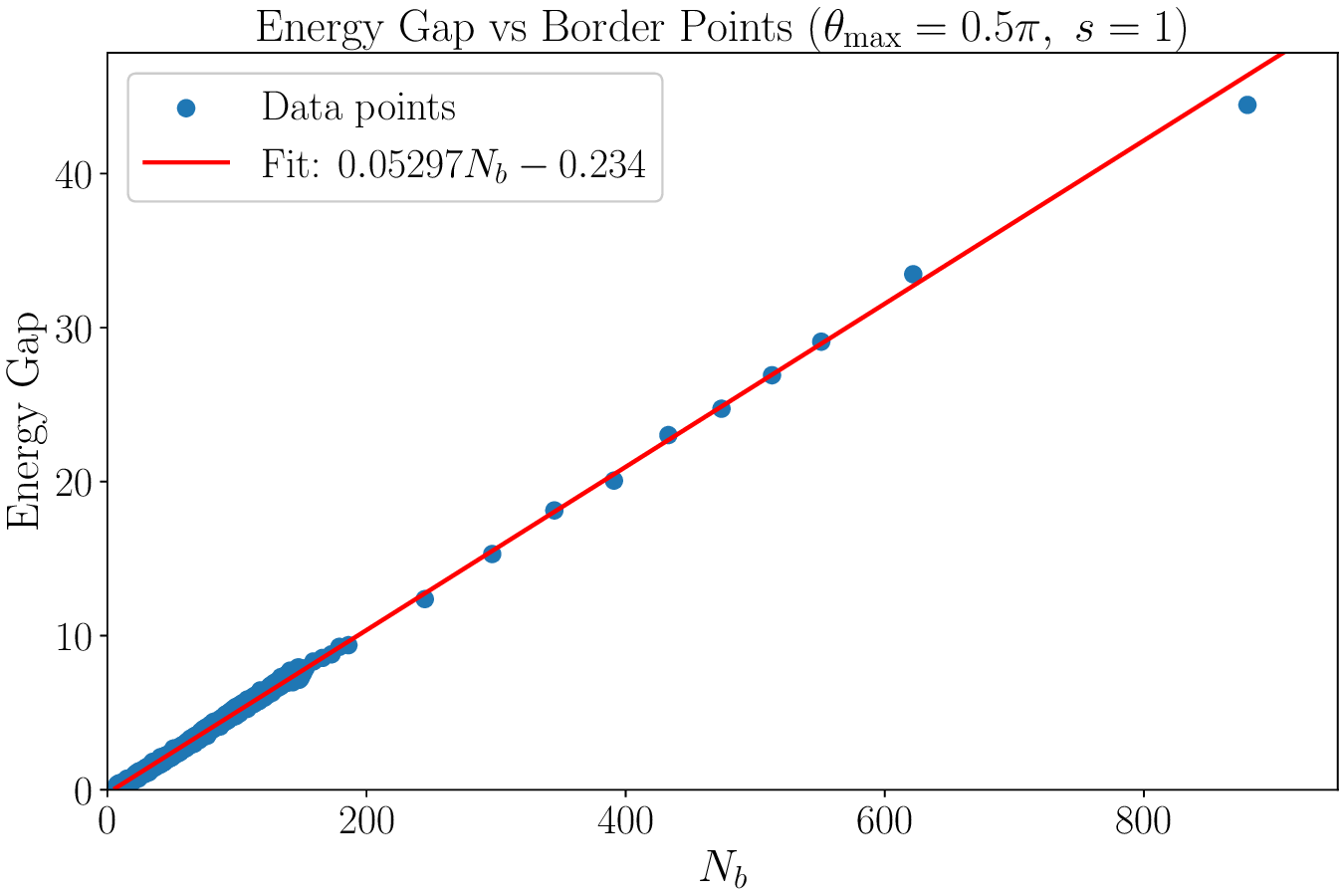}
\caption{Energy gap as a function of $N_b$ for configurations of points on the hemisphere.}
\label{Fig_gap_05}
\end{center}
\end{figure}

In Fig.~\ref{Fig_gap_05} we display the energy gap for the configurations of points on the hemisphere, for the cases that we have calculated. 
This gap is defined as the difference between the lowest individual energy of an internal charge and the largest individual energy of a border charge.
The curve in this plot correspond to a two--parameters linear fit of the data  $E_{\rm gap} \approx 0.05312 N_b- 0.2478$.

\section{Conclusions}
\label{sec:concl}

We have studied Thomson problem on a  spherical cap, amounting to finding the configurations of $N$  charges confined to the cap, mutually repelling via the Coulomb force. While this problem had never been studied before for arbitrary angular widths, it contains two well--known problems as limit cases: Thomson problem itself (when the cap reduces to the whole sphere) and Thomson problem on the disk, when the angular width is made arbitrarily small, while maintaining the area constant (the relevant bibliography on these two problems is cited in the Introduction).

\bigskip

We briefly state the main findings:
\begin{itemize}
\item we have implemented an efficient computational scheme to find low--energy configurations  of $N$ points on the spherical cap and we have carried out multiple numerical simulations for caps of different widths ($\theta_{\rm max}/\pi =0.1,  \dots, 0.9$): for the special case of the hemisphere
we have produced results for $10 \leq N \leq 737$ and for selected values of $N$ ranging up to $N=10000$;

\item we have compared the numerical results for the discrete model with the exact result for the continuum, which had been found long time ago by Lord Kelvin: for $N$ sufficiently large  it is observed that the discrete results approach the continuum limit; moreover, we have estimated the correlation energy for the system, using the {\sl local density approximation} and found an excellent agreement with the estimates obtained from the numerical results; 

\item  we have worked out an approximate formula for the number of charges deposited on the border using the Lord Kelvin's solution for the  continuum and we have found that $N_{border} \propto N^{2/3}$, as for the flat disk,  regardless of the width of the cap. Our numerical results confirm these findings. It is natural to ask oneself whether this behavior could be valid for charges mutually repelling via the Coulomb force, on  more general surfaces with border.

\item we have looked at the Voronoi diagrams for the configurations that we have produced: for a small angular width, it is seen that the border is populated with roughly equal amount of pentagonal and square cells, leading to a border topological charge proportional to $N_{border}$ (and therefore $\propto N^{2/3}$); as the angular width increases the topological charge on the border is decreasing and as a result a larger number of defects enters the bulk; 

\item while Euler's theorem constrains the total topological charge, it does not put restrictions on the number of defects in a configuration: for this reason, we have used our numerical  results to estimate the growth of the number of internal defects (defects that are not on the border) as $N$ increases;

\item for $N$ sufficiently large, the density of the discrete configurations is seen to nicely approach the continuum density, originally found by Lord Kelvin;

\bigskip

\item the Voronoi cells display a larger strain close to the regions where topological defects are present;

\end{itemize}

In future work we plan to extend this study to a larger class of  (long and short range) potentials and possibly to a larger class of domains with both border and curvature.  
The effect of the potentials and of the characteristics of the domain on the 
, in particular in relation to the question of how curvature and border affect 
the patterns of defects for the different potentials.

\section*{Acknowledgements}
I am grateful to Professor Alfredo Aranda for useful discussions. My research is supported by Sistema Nacional de Investigadoras e  Investigadores (M\'exico). 

\section*{Data availability}
A graphic catalog of the configurations obtained in this work can be found at Zenodo. Data will be made available upon request.

\end{document}